\documentclass[a4paper,10pt]{article}

\usepackage[latin1]{inputenc}
\usepackage[T1]{fontenc}
\usepackage{ae,aecompl}
\usepackage{epsfig}
\usepackage{amsmath}
\usepackage{amsfonts}
\usepackage{relsize}
\usepackage{amssymb}
\usepackage{a4wide}
\usepackage{doublespace}
\usepackage{amssymb}
\usepackage{bm}

\newcommand{\diag}{{\rm diag}}
\newtheorem{hyp}{Assumption}
\newtheorem{prop}{Proposition}
\newtheorem{definition}{Definition}
\newcommand{\eqv}[1]{\mathop{\sim}\limits_{#1}}
\newcommand{\LLR}{\textrm{\em LLR}}
\newcommand{\tr}{{\rm Tr}}
\newcommand{\mat}[1]{\mathbf{#1}}
\newcommand{\matgr}[1]{\bm{#1}}
\renewcommand{\H}{\mathcal{H}}
\renewcommand{\mod}[2]{#1\textrm{~mod~}#2}
\newcommand{\C}[2]{\binom{#1}{#2}}
\newcommand{\cqfd}{\ifmmode\sqw\else{\ifhmode\unskip\fi\nobreak\hfil \penalty50\hskip1em\null\nobreak\hfil\sqw \parfillskip=0pt\finalhyphendemerits=0\endgraf}\fi} 

\title{Space-time coding techniques with bit-interleaved coded modulations for MIMO block-fading channels\\
\vskip -10pt
\begin{normalsize}\textit{Submitted to IEEE Trans. on Information Theory}\end{normalsize}\\
\vskip -30pt
\begin{normalsize}\textit{Submission: January 2006   -   First review: June 2007}\end{normalsize}\\
}
\author{Nicolas Gresset, Loïc Brunel and Joseph Boutros\\
\smaller{ENST Paris, 46 rue Barrault, 75013 Paris, France}\\
\smaller{Mitsubishi Electric ITE-TCL, 1 allée de Beaulieu, 35700 Rennes, France}\\
\smaller{gresset@tcl.ite.mee.com, brunel@tcl.ite.mee.com, boutros@ieee.org}}
\date{}

\begin{document}

\maketitle

\begin{abstract}
The space-time bit-interleaved coded modulation (ST-BICM) is an efficient
technique to obtain high diversity and coding gain on a block-fading
MIMO channel. Its maximum-likelihood (ML) performance is computed
under ideal interleaving conditions, which enables a global optimization
taking into account channel coding.
Thanks to a diversity upperbound derived from the Singleton bound,
an appropriate choice of the time dimension of the space-time coding 
is possible, which maximizes diversity while minimizing complexity.
Based on the analysis, an optimized interleaver and a set of linear precoders,
called dispersive nucleo algebraic (DNA) precoders are proposed.
The proposed precoders have good performance with respect to the state of the art 
and exist for any number of transmit antennas and any time dimension.
With turbo codes, they exhibit a frame error rate 
which does not increase with frame length.
\noindent
\begin{center}
{\bf Index terms}\\
Multiple antenna channels, bit-interleaved coded modulation, space-time coding, Singleton~bound, interleaving
\end{center}
\end{abstract}

\section{Introduction}
The wide panel of today's wireless transmission contexts 
makes implausible the existence of a miraculous universal
solution which always exhibits good performance with low complexity.  
Different scenarios (indoor, outdoor with low velocity, outdoor with high
velocity)
correspond to different amounts of time and frequency diversity.
The success of the multi-carrier modulation as a solution
for future wireless systems is in part due to
the low receiver complexity even over large frequency bands.
In this paper, we focus on an indoor environment and design 
a system approaching the optimal performance taught by  
information theory. 
In a wireless indoor environment, both time and frequency diversities
may be poor due to small terminal velocity and possibly very short channel
impulse response.
These particularly tricky low-diversity channels are modelled as
block-fading channels.
Over low-diversity multiple-input multiple-output (MIMO) channels,
space-time coding techniques often enable
transmission with improved data rate and diversity,  
within a limit given by the rank of the MIMO system 
\cite{Alamouti1998}\cite{Elgamal2003}\cite{Elgamal2003-2}\cite{Foschini1996}.
These open-loop schemes only require the knowledge of the channel long-term statistics.
Besides, closed-loop techniques such as beamforming take benefit
from a short-term channel knowledge to improve the performance/complexity trade-off at the cost 
of additional signalling overhead.
As a first step in providing increased data rates 
in future generations of indoor wireless local access networks (WLANs),
we study how to appropriately choose
the channel coding, the channel interleaving and the space-time coding. 
\\

For frame sizes of practical interest, coded modulations have to be considered
since space-time codes employed with uncoded modulations 
exhibit a frame error rate (FER), which is dramatically degraded \cite[Annex A]{Gresset2004-Thesis}.
Thus, we focus on the bit-interleaved coded modulation (BICM) structure, 
which is the concatenation of a channel encoder,
an interleaver and a modulator. The analysis of the BICM maximum-likelihood (ML) performance 
is tractable and eases the coded modulation design.
Furthermore, thanks to the interleaver, iterative processing at the receiver
achieves quasi-ML performance with reduced complexity.
On a MIMO channel, the BICM may be concatenated with a simple full-rate space division multiplexing
scheme (SDM) \cite{Foschini1996}. In this paper, we improve performance
of this space-time BICM (ST-BICM) by replacing the SDM by a more efficient
full-rate linear space-time code: a linear precoding 
or equivalently a space-time spreading.
Linear precoding is performed by multiplying the complex multiple-antenna signal
by a square complex space-time matrix.
The space-time matrix enhances the diversity by mixing the symbols
of different time periods and antennas together.
\\

The choice of the ST-BICM structure may also be explained as follows:
We aim at optimizing a full-rate space-time code based on linear precoding, 
taking into account the structure of the whole transmitter, which inevitably
includes an error correction code, an interleaver and a symbol mapper.
Usually, space-time codes are designed independently from the other elements
of the transmitter. However, frames of bits are linked through the error correction
code and optimizing the space-time code taking into account the whole transmitter
is equivalent to optimizing a BICM concatenated with a space-time code, i.e., an ST-BICM.\\

On an ergodic channel, the achieved diversity order is equal to the code minimum
distance multiplied by the number of receive antennas. 
In most cases, the minimum distance is high enough
and increasing diversity through linear precoding does not bring much improvement.
For a block-fading channel, the diversity is upperbounded by the number
of channel realizations in a codeword multiplied by the number of transmit antennas
and the number of receive antennas. Using the Singleton bound, we will exhibit 
an additional upperbound on the diversity order, which may be very limiting 
without precoding.
Hence, in this paper, we will study ST-BICM with linear precoding,
focusing on the block-fading channel and optimize the linear precoding using the ST-BICM ML performance
in order to achieve full-diversity and maximum coding gain. 
First, we derive the coding gain of an ideal ST-BICM.
It is related to the notion of Shannon code
and sphere-hardening \cite{Shannon1959}.
Indeed, the ideal Shannon code for additive white gaussian noise (AWGN) channels is located
near a sphere, called the Shannon's sphere.
Thanks to the interleaver, the squared Euclidean norm of BICM codewords has low variance,
which implies that codewords lie close to the Shannon's sphere.
The BICM may be seen as a quantization of this sphere, which should be as uniform as possible 
to maximise the size of Voronoi regions.
On MIMO fading channels, the Shannon's sphere becomes a Shannon's ellipsoid \cite{Fozunbal2003}
and BICM codewords are randomly located close to the ellipsoid.
We show that the ideal BICM configuration
maximizes the Voronoi region volume whatever the channel realization.
We present a practical system that approaches the ideal BICM configuration 
including the so-called dispersive nucleo algebraic (DNA) precoder
and compare its performance to the ideal ST-BICM performance.
The DNA precoder exists for any numbers of space
and time dimensions. We finally design a practical interleaver, which approximates the
ideal interleaving conditions.\\

The paper is organized as follows: In section \ref{sec:system}, the ST-BICM transmitter
and the associated iterative receiver are presented. In section \ref{sec:theo_perf},
we derive the analytical ML performance under ideal interleaving assumption for an ergodic
channel without precoding and a block-fading channel with and without precoding.
Using the Singleton bound, we show in section \ref{sec:singleton_bound} that ideal interleaving
conditions cannot be achieved on a block-fading channel 
with any kind of parameters and that linear precoding may be mandatory in some configurations.
Section~\ref{sec:linear-prec} describes the linear precoding optimization for a block-fading channel
and section~\ref{sec:interleaver} the interleaver design for convolutional codes
and its application to turbo codes.
Finally, simulation results
are presented in section~\ref{sec:results}, which confirm the behavior which was
expected from the analytical study. Furthermore, they show the good performance 
of the DNA precoders and the advantage of using turbo codes to get a non-increasing FER
when the frame size increases.

\section{System model and notations \label{sec:system}}

\subsection{Transmitter scheme}
\begin{figure}[!ht]
\centering
\includegraphics[width=0.9\columnwidth]{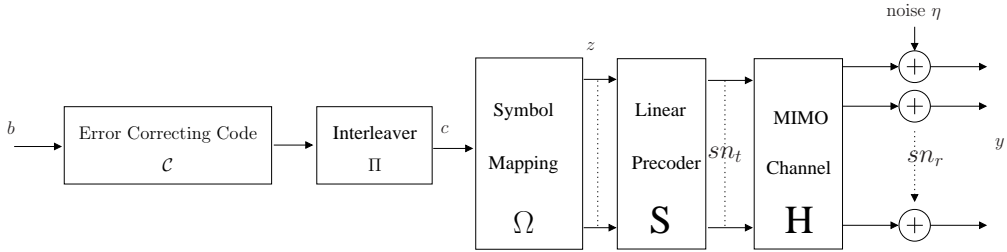}
\caption{Bit-interleaved coded modulation transmitter and multiple antenna channel model.}
\label{fig:system_model}
\end{figure}

The transmitter scheme is built from the following fundamental block concatenation:
A binary error-correcting code $\mathcal{C}$
followed by a deterministic interleaver $\Pi$, a symbol mapper (e.g., for a quadrature amplitude modulation (QAM)),
a full-rate space-time spreader $\mat{S}$ (i.e., a linear precoder)
and a set of $n_t$ transmit antennas. Fig. \ref{fig:system_model} illustrates
the BICM transmitter structure.\\

Without loss of generality, we assume that the error correcting code $\mathcal{C}$
is a convolutional code with rate $R_\mathcal{C}$.
The encoder associates with the input information word $\mat b$ the codeword $\mat c\in\mathcal{C}$.
Sequence $\mat b$ (resp. $\mat c$) has length $K_\mathcal{C}L_\mathcal{C}$ (resp. $N_\mathcal{C}L_\mathcal{C}$) bits,
where $L_\mathcal{C}$ is the codeword length in trellis branches.
The interleaver $\Pi$, which scrambles the $L_\mathcal{C}N_\mathcal{C}$ coded bits, is a crucial
function in the BICM structure, as it allows the receiver to perform iterative joint detection
and decoding. Indeed, it ensures independence between extrinsic and a priori probabilities,
in both the detector and the decoder.
Furthermore, when maximum-likelihood (ML) decoding is tractable,
interleaving prevents erroneous bits of a same error event from interfering to each other
in the same precoded symbol.
The interleaver $\Pi$ may be pseudo-random (PR) or semi-deterministic with some deterministic constraints
as described in section \ref{sec:interleaver}.
In the symbol mapper, $m$ consecutive interleaved coded bits are mapped together onto a modulation symbol,
according to a bijection between bit vectors and modulation symbols called mapping or labeling.
The number of modulation symbols is equal to $M=2^m$.
For each channel use, i.e., in each time period, the mapper reads $mn_t$
coded bits and generates $n_t$ modulation symbols. To make the reading easier,
the obtained $n_t$-dimensional constellation $\Omega$ will denote both the set of symbols 
and the set of binary labelings.
All along this paper, we will consider QAM modulations 
as they achieve a good compromise between spectral efficiency (in bits/s/Hz or bits/dim) and performance. 
Moreover, with QAM modulation, the system is easily modeled using a lattice constellation structure \cite{Damen2000},
which gives access to the lattice theory toolbox, both for transmitter and receiver optimizations.
We assume that the QAM modulation has unit energy.
The linear precoder $\mat{S}$ spreads the QAM symbols over $s$ time periods. It converts
the $n_t \times n_r$ vector channel into an $N_t \times N_r$ vector channel,
where $N_t=n_ts$ and $N_r=n_rs$.
The $N_t \times N_t$ matrix $\mat{S}$ multiplies a vector of $N_t$ QAM symbols 
$\mat{z}_i=({z}_{i,1}, {z}_{i,2}, \ldots, {z}_{i,N_t})$ at the mapper output, generating $N_t$ symbols 
to be transmitted during $s$ time periods. Vector $\mat{z}_i$ is the $i^{\textrm{th}}$ vector to be precoded.
The precoder $\mat{S}$ spreads the transmitted symbols over a higher number of channel states
to exploit diversity. $\mat{S}$ is normalized as follows:
\begin{equation}
\sum_{u=1}^{N_t}\sum_{v=1}^{N_t}\mat{S}_{u,v}^2=N_t
\end{equation}
In this paper, we assume a block-fading channel with $n_c$ distinct
channel realizations during a codeword. We denote $n_s$ the 
number of distinct channel realizations during a precoded symbol.
To simplify notations, we assume that $n_s$ divides $n_c$.
We will call {\em channel state} the $1\times n_r$
SIMO channel associated with one of the $n_t$ transmit antennas and one of the $n_c$
channel realizations.
The channel experienced by precoded symbol $i$ is represented by
a $N_t \times N_r$ block-diagonal matrix $\mat{H}_i$ with $s$ blocks of size $n_{t} \times n_r$.
During one precoded symbol, we assume that each of the $n_s$ channel realizations is repeated
$s/n_s$ times. The $\mat{H}_i$ matrix is organized as follows:
\begin{equation} \label{eq:Hk_general}
\mat{H}_i=\diag\left(\mat{H}_{i}^{[1]},\ldots,\mat{H}_{i}^{[1]},\mat{H}_{i}^{[2]},\ldots,\mat{H}_{i}^{[2]},
\ldots,\mat{H}_{i}^{[n_s]}\ldots,\mat{H}_{i}^{[n_s]}\right)
\end{equation}
where $\mat{H}_{i}^{[t]}$ denotes the $n_{t} \times n_r$ complex matrix
representing the $t$-th channel realization
experienced by the $i$-th precoded symbol. $\mat{H}_{i}^{[t]}$ is repeated $s/n_s$ times.
Elements of $\mat{H}_{i}^{[t]}$ are independent complex Gaussian variables with
zero mean and unit variance.
Let $\H$ denote the set of channel realizations observed during the transmission of a codeword.
Thanks to the extended channel matrix, we write the channel input-output relation as:
\begin{equation}\label{eq:inputoutput}
\mat{y}_i=\mat{x}_i+\matgr{\eta}_i=\mat{z}_i\mat{S}\mat{H}_i+\matgr{\eta}_i
\end{equation}
where $\mat{y}_i\in \mathbb{C}^{N_r}$ and each receive antenna is perturbed by an additive white complex Gaussian
noise $\eta_{i,j}$, $j=1\ldots N_r$, with zero mean and variance $2N_{0}$.
We define the signal-to-noise ratio $E_b/N_0$, where $E_b$ is the total energy of an information bit at the receiver.
Thanks to linear precoding, the $n_t\times n_r$ MIMO $n_c$-block-fading channel 
is converted into an $N_t\times N_r$ MIMO $N_c$-block-fading channel where $N_c=n_c/n_s$.
If $n_s=1$, the precoder experiences a quasi-static $n_t\times n_r$ MIMO channel.
In the following, index $i$ will be omitted if a single precoded symbol is considered and
{\em precoding time period} will refer to a transmission over $\mat{S}\mat{H}$, i.e., over
$s$ time periods.\\

The concatenation of the binary error correcting code $\mathcal{C}$, the interleaver $\Pi$, the mapper $\Omega$, the
linear precoder $\mat{S}$ and the channel describes a global Euclidean code $\mathcal{C}_E$ which converts 
$L_\mathcal{C}K_\mathcal{C}$ information bits into a complex $L_\mathcal{C}N_\mathcal{C}/m$-dimensional point.

\subsection{Iterative receiver scheme}
\begin{figure}[!ht]
\centering
\includegraphics[width=0.9\columnwidth]{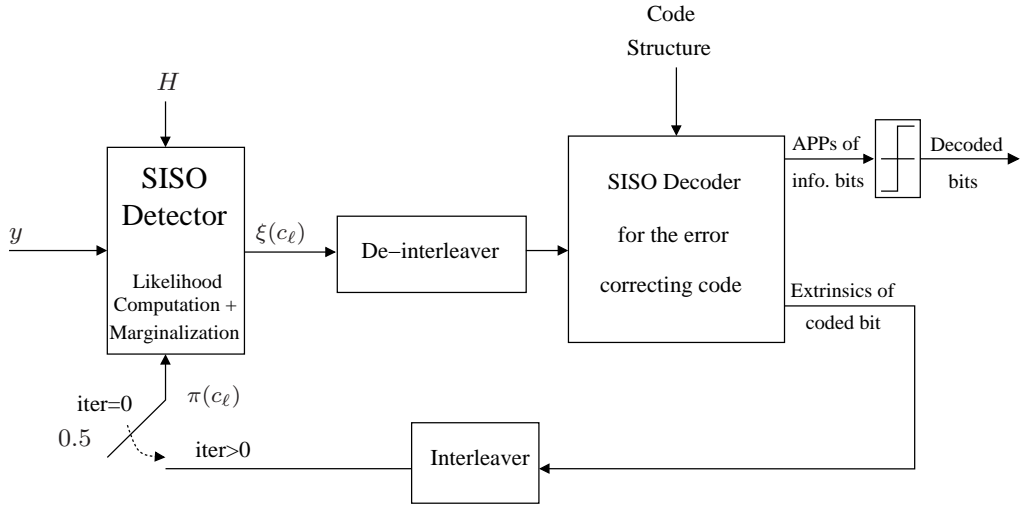}
\caption{Iterative APP detection and decoding receiver.}
\label{fig:detect_decode}
\end{figure}

An ideal BICM receiver would directly perform an ML decoding on the set $\mathcal{C}_E$ of transmitted codewords. 
However, it requires an exhaustive  search among the $2^{K_\mathcal{C}L_\mathcal{C}}$ codewords, 
which is intractable.
All existing receivers use the concatenated structure of the BICM to split the reception into several steps. 
In this paper, we assume perfect synchronization and channel estimation.
Thus, the receiver, as depicted on Fig. \ref{fig:detect_decode}, is divided in two
main elements: a soft-input soft-output (SISO) APP QAM detector, which acts as a soft-output equalizer for both the space-time
spreader and the MIMO channel, converting the received point $\mat{y}$ into information on the coded bits in 
the estimated coded sequence $\hat{\mat{c}}$,
and a SISO decoder for $\mathcal{C}$, improving the information on coded bits
and estimating the information bit sequence $\hat{\mat{b}}$.
The depicted iterative joint detection and decoding process is based on the exchange of soft values 
between these two elements. 
The SISO detector computes extrinsic probabilities $\xi (c_{\ell })$ on coded bits
thanks to the conditional likelihoods $p(\mat{y}_i/\mat{z})$ and the {\em a priori} probabilities $\pi (c_{\ell })$
fed back from the SISO decoder:
\begin{equation}\label{eq:marginalization}
\xi (c_{\ell})=\frac{\sum _{\mat{z}'\in \Omega (c_{\ell}=1)}\left[\left(e^{-\frac{\left\Vert \mat{y}_i-\mat{z}'\mat{S}\mat{H}_i\right\Vert ^{2}}{2N_0}}\right)\prod _{r\neq \ell }\pi (c_{r})\right]}{\sum _{\mat{z}\in \Omega }\left[\left(e^{-\frac{\left\Vert \mat{y}_i-\mat{z}\mat{S}\mat{H}_i\right\Vert ^{2}}{2N_0}}\right)\prod _{r\neq \ell }\pi (c_{r})\right]}
\end{equation}
\noindent
where $\Omega $ is the Cartesian product $(M\textrm{-QAM})^{N_{t}}$,
i.e., the set of all vectors $\mat{z}$ generated by the QAM mapper, $|\Omega |=2^{mN_{t}}$.
The subset $\Omega (c_{\ell }=1)$, for $\ell = 0, 1, \ldots, mN_ t-1$, is restricted to the vectors $\mat{z}$
in which the $\ell$-th coded bit is equal to $1$.
The detector independently computes the soft outputs for each precoding time period.
At the first iteration, no a priori information is available at the detector input. 
Through the iterations, the a priori probability on constellation points computed from the probabilities
fed back by the SISO decoder becomes more and more accurate. Ideal convergence is achieved 
when a priori probabilities provided by the decoder are perfect, i.e., equal to $0$ or $1$.
The decoder uses a forward-backward algorithm \cite{Bahl1974}, which computes the exact extrinsic probability
using the trellis structure of the code.

\section{Theoretical performance for ideally interleaved BICM \label{sec:theo_perf}}

Heavy work has been made to estimate the frame or bit error rate of the
BICM with ML decoding,
in particular using Gaussian approximations or numerical integrations \cite{biglieri2002},
but a closed-form expression of the pairwise error probability had not been derived yet.
This section first describes an accurate computation of bit and frame error rates
of BICM ML performance over ergodic MIMO channel with ideal interleaving
and without precoding.
A more detailed description of the derivation may be found in \cite{Gresset2004-IT}
and \cite{Gresset2004-Thesis}.
Under ideal interleaving condition, we are able to derive a closed form expression
of the probability density of the log likelihood ratio (LLR) at the output of the detector 
and then a closed form expression of the pairwise error probability at the output of the decoder.
It is then straightforward to use well-known techniques to estimate the bit or frame error rate 
of a coded modulation from pairwise error probability.
This subject has been extensively discussed for coded modulations over AWGN channels.
Examples are the union bound on the transfer function of a convolutional code
and the more accurate tangential sphere bound \cite{Poltyrev1996} for spherical constellations.

Subsequently, we extend the study to the block-fading MIMO channel when linear precoding 
is used at the transmitter.
Note that the method is also valid for correlated MIMO channels.
We extract from the bit error rate expression some design
criteria on BICM precoder, interleaver, and error correcting code.

\subsection{Ideal interleaving condition}
The evaluation of the bit error rate (BER) or frame error rate (FER) of a 
coded modulation
is usually based on the derivation of an upper bound on the actual performance
obtained by a balanced summation of pairwise error probabilities.
Each pairwise error probability involves the Euclidean distance between two 
codewords
with a Hamming distance $w$.\\

With an $n_t\times n_r$ MIMO block-fading channel with $n_c$ 
blocks, the minimum diversity recovered at the detector output, and thus at
the decoder output, is always equal to the reception diversity  $n_r$.
Let us consider an error event with $w$ erroneous bits. Assume that the 
maximum diversity
order is $\Upsilon_{max}$. If $w\geq \Upsilon_{max}/n_r$, we achieve full 
diversity if each of the $\Upsilon_{max}/n_r$ independent fading random variables
is experienced by at least one bit among $w$. 
In a precoding time period $k$ in which
at least an erroneous bit is transmitted, the transmitted and competing 
points are called $\mat{x}_k=\mat{z}_k\mat{S}\mat{H}_k$ and $\mat{x}'_k=\mat{z}'_k\mat{S}\mat{H}_k$.
When performing ML decoding or APP detection, we are interested 
in the equivalent Binary Shift Keying (BSK) modulation defined by the two points $\mat{x}_k$ and $\mat{x}'_k$. 
The vector $(\mat{z}_k-\mat{z}'_k)\mat{S}\mat{H}_k$ has $sn_r$ circular symmetric Gaussian components.
Thus, whatever the number of erroneous bits on a precoding
time period, the obtained diversity is limited to $sn_r$. 
Having several erroneous bits per precoding time period is useless.
On the contrary, if the erroneous bits are located on
different precoding time periods and experience different fading random
variables, 
a higher diversity is achieved. This is what we call the non-interference
property.
Furthermore, we will see in section \ref{sec:linear-prec}
that an equi-distribution of erroneous bits on channel states is required
to achieve a maximum coding gain. We call it the equi-distribution property.
The ideal interleaver is defined as follows:
\begin{definition} \label{def:ideal_interleaving}(Ideal Interleaving)
For any pair of codewords with $w$ different bits at positions $i_1, \ldots, 
i_k, \ldots, i_w$,
an ideal interleaver allocates the bits to transmitted symbols as follows:
\begin{itemize}
\item Non-interference property: $\forall i_k, i_{k'}$, bits at positions $i_k$ and $i_{k'}$ are 
transmitted on different precoding time periods,
\item Equi-distribution property: the bits at positions $i_1, \ldots, i_k, \ldots, i_w$ are as 
equiprobably distributed over
all channel states as allowed by $w$.
\end{itemize}
\end{definition}

In practice, such an interleaver does not always exist. We will see in the 
following that the
Singleton bound gives an existence condition of the ideal interleaver. In 
section
\ref{sec:interleaver}, we present optimized interleavers that approach the 
ideal condition.

\subsection{Exact pairwise error probability for ergodic channels without precoding}

In \cite{Gresset2004-IT}, we have established a closed form expression for the conditional pairwise error probability
on ergodic MIMO channels under ML decoding of the BICM and ideal channel interleaving.
The mathematical derivation in this subsection follows \cite{Gresset2004-IT}.
Transmitted symbols are not precoded: $s=1$, $\mat{S}=\mat{I}_{n_t}$ the $n_t \times n_t$ identity matrix.
Thus, (\ref{eq:Hk_general}) reduces to $\mat{H}_k = \mat{H}_{k}^{[1]}$.
Consider the pairwise error probability that a codeword $\mat{c}\in\mathcal{C}$ is transmitted and
a codeword $\mat{c}'\in\mathcal{C}$ is decoded.
The $w$ different bits between the two codewords are transmitted in $w$ different time periods,
complementing one bit in the mapping of one of the $n_t$ QAM symbols.
The transmitted noiseless vectors corresponding to the two codewords $(\mat{c},\mat{c}')$ only differ in $w$ positions.
Let us define $\mat{Z}=(\mat{z}_1,\dots,\mat{z}_w)$ and $\mat{Z}'=(\mat{z}'_1,\dots,\mat{z}'_w)$ the $wn_t$-dimensional vectors corresponding
to these positions and $\mat{X}=(\mat{x}_1,\dots,\mat{x}_w)$ and $\mat{X}'=(\mat{x}'_1,\dots,\mat{x}'_w)$ the $wn_r$-dimensional
vectors corresponding to $\mat{Z}$
and $\mat{Z}'$ and filtered by the channel matrix $\mat{H}_{k}$.

We define $d_k=\|\mat{z}_k-\mat{z}'_k\|$. The Euclidean distance $\|\mat{X}-\mat{X}'\|$ depends on both the set of distances $\{d_1,\ldots,d_k,\ldots,d_w\}$
and the set of channel realizations $\H$.
Let $D$ denote the set of all Euclidean distances obtained by flipping one bit in the constellation $\Omega$.
Define the set $\Delta=\{\delta_1,\ldots,\delta_{n_d}\} \subset D$ with distinct elements from the sequence
$(d_1, d_2, \ldots, d_w) \in \Delta^w \subset D^w$, i.e., the Euclidean distance $d_k$
takes its values from the set $\Delta$. Obviously, $n_d=|\Delta| \le |D|$.
Let the integer $\lambda_k$ denote the frequency of $\delta_k$ in the sequence
$(d_1, d_2, \ldots, d_w)$, $\sum_{n=1}^{n_d}\lambda_n=w$ and $\Lambda=\left\{\lambda_1,\ldots,\lambda_{n_d}\right\}$.
The pairwise error probability
conditioned on the channel realization set $\H$ and the Hamming weight $w$ is expressed as
\begin{equation}
P_{w,\H}(\mat{c}\rightarrow \mat{c}')=P_{w,\H}(\mat{X}\rightarrow \mat{X}')=P\left(\sum_{k=1}^{w} \LLR_{k}<0\right) \label{eq:cond_pairwise}
\end{equation}
where $\LLR_{k}$ is the $k$-th LLR, corresponding to the $k$-th error position, and is equal to
\begin{equation}
\LLR_{k}=\frac{\|\mat{y}_k-\mat{x}'_k\|^2-\|\mat{y}_k-\mat{x}_k\|^2}{2N_0}\sim \mathcal{N}\left(\frac{R_k}{2N_0},\frac{R_k}{N_0}\right)
\end{equation}
$R_k=\left\|(\mat{z}_k-\mat{z}'_k)\mat{H}_{k}\right\|^2$  has a chi-square distribution of order $2n_r$.
Averaging over $\H$, we calculate the characteristic function of 
$E_\H\left[\sum_{k=1}^w \LLR_k\right]$:
\begin{equation}
\psi(j\nu)
=\left(\prod_{k=1}^{w}\left(\frac{-d_k^2}{2N_0}\right)^{-n_r}\right)
\left(\prod_{n=-n_d,n\neq0}^{n_d}[j\nu+\beta_n]^{-n_r\lambda_{|n|}}\right)
\end{equation}
where
\begin{equation}
\left\{
\begin{array}{ll}
n>0,& \beta_n=\frac{1}{2}\left(1+\sqrt{1+\frac{8N_0}{\delta_n^2}}\right)\\
n<0,& \beta_n=\frac{1}{2}\left(1-\sqrt{1+\frac{8N_0}{\delta_n^2}}\right)\\
\end{array}
\right.
\end{equation}
Applying a partial fraction expansion, we obtain the expression of the pairwise error probability:
\begin{equation}\label{eq:closed_pairwise}
P_w(\mat{X}\rightarrow \mat{X}')=P_w(\Delta,\Lambda)=\prod_{k=1}^{w}\left(-\frac{2N_0}{d_k^2}\right)^{n_r}
\sum_{n=1}^{n_d}\sum_{i=1}^{n_r\lambda_n}\frac{\alpha_{n,i}}{\left(\frac{1}{2}+\frac{1}{2}\sqrt{1+\frac{8N_0}{\delta_n^2}}\right)^i}
\end{equation}
where the coefficients $\alpha_{n,i}$ are given by an identification
of the coefficients of two series expansions in $\epsilon$ as in \cite{Gresset2004-IT}.

We compute the asymptotic expression when the noise level is low. Indeed, the coding gain and
diversity are measured for high signal-to-noise ratios, where the performance has a linear asymptote
on logarithmic scales.
\begin{equation}
P_w(\Delta,\Lambda)\eqv{N_{0}\rightarrow 0}
\C{2n_{r}w-1}{n_rw}\prod_{k=1}^{w}\left(\frac{2N_0}{d_k^2}\right)^{n_r}
=\C{2n_{r}w-1}{n_rw}\left(\frac{2N_0}{\mathcal{G}_{ergo}(\Delta,\Lambda)}\right)^{wn_r}
\end{equation}
with $\C{n}{k}=n!/(k!(n-k)!)$.
The diversity associated with the considered pairs of Hamming weight $w$ is the
exponent of $2N_0$, equal to $wn_r$.
We define the coding gain
or coding advantage as the coefficient dividing $2N_0$, i.e.,
\begin{equation}
\mathcal{G}_{ergo}(\Delta,\Lambda)
=\left(\prod_{k=1}^{w}d_k^{2}\right)^{1/w}
\end{equation}
All sequences $(d_1,\ldots,d_w)$ corresponding to the same pair
$\left(\Delta,\Lambda\right)$ yield the same pairwise error probability.
By averaging over all possible pairs $(\mat{c},\mat{c}')$
or equivalently over all sets of distances $D^w$, we obtain
$P_w=E_{D^w}\left[P_w(\Delta,\Lambda)\right]$, the conditional probability
that an error event of Hamming weight $w$ occurs.
From this pairwise error probability, it is easy to estimate the FER or BER of the BICM 
with ideal interleaving thanks to a classical union bound on the weight enumeration
function of the error correcting code. 
Moreover, we may derive a design criterion of the BICM
from the coding gain $\mathcal{G}_{ergo}(\Delta,\Lambda)$ expression. 
In the following, we derive the coding gain for block-fading channels 
and linear precoding in order to obtain the ML design criterion of the ST-BICM.

\subsection{Exact pairwise error probability for MIMO block fading channels without precoding}
We assume that Definition 1 is satisfied.
For a block-fading channel with $n_c$ independent realizations in a frame, 
the decision variable between $\mat{X}$ and $\mat{X}'$ is still given by (\ref{eq:cond_pairwise}).
However, the involved channel matrices are not independent as for an ergodic channel.
The conditions of independence are the following:
\begin{itemize}
\item If two LLR random variables depend on two different channel realizations, they are independent.
\item If two LLR random variables depend on the same channel realization but on different
transmit antennas, the random variables are independent. 
\end{itemize}

The maximum number of independent LLR variables is $n_cn_t$, the
transmit diversity order. 
We choose the error correcting code so that $w \geq n_tn_c$.
We now group the $w$ random variables LLR into $\min(n_tn_c,w)=n_tn_c$ independent blocks.
Let $\LLR_{k,l,i}$ be the $i$-th log-likelihood ratio corresponding to the BSK transmission
on the $l$-th antenna of the $k$-th block,
$k=1 \ldots n_c$, $l=1 \ldots n_t$ and $i = 1 \ldots \kappa_{k,l}$, where
$\kappa_{k,l}$ is the number of bits transmitted on the $l$-th antenna of the $k$-th block.
We have $\sum_{k=1}^{n_c}\sum_{l=1}^{n_t}\kappa_{k,l}=w$.
Finally,  $\LLR$ is the sum of the $n_tn_c$ independent random variables $\LLR_{k,l}=\sum_{i=1}^{\kappa_{k,l}}\LLR_{k,l,i}$.
Let $d_{k,l,i}$ denote the distance associated with $\LLR_{k,l,i}$, and define
$\gamma_{k,l}^2=\sum_{i=1}^{\kappa_{k,l}}d_{k,l,i}^2$ the distance associated with $\LLR_{k,l}$.
We have
\begin{equation}
\LLR_{k,l}\sim
\mathcal{N}\left(\frac{R_{k,l}}{2N_0},
\frac{R_{k,l}}{N_0}\right)
\end{equation}
where $R_{k,l}=\gamma_{k,l}^2\|\mat{H}_k(l)\|^2$ and $\mat{H}_k(l)$ is the $l$-th row of $\mat{H}_k$.
For all $i$, $\LLR_{k,l,i}$ are transmitted over the equivalent $1\times n_r$ SIMO channel defined by $\mat{H}_k(l)$,
which is chi-square distributed with degree $2n_r$.
The $\LLR_{k,l}$ variables are transmitted on independent channel states, as for the ergodic channel case,
we directly apply (\ref{eq:closed_pairwise}) and obtain the conditional pairwise error
probability closed-form expression
\begin{equation}
P_w(\mat{X}\rightarrow \mat{X}')=P_w(\Delta,\Lambda)=\prod_{k=1}^{n_c}\prod_{l=1}^{n_t}
\left(-\frac{2N_0}{\gamma_{k,l}^2}\right)^{n_r}
\sum_{n=1}^{n_d}\sum_{i=1}^{n_r\lambda_n}\frac{\alpha_{n,i}}{\left(\frac{1}{2}+\frac{1}{2}\sqrt{1+\frac{8N_0}{\delta_n^2}}\right)^i}
\end{equation}
where $\delta_n\in\Delta$ and $(\Delta,\Lambda)$ is the pair of sets
representing the sequence
$\left(\gamma_{1,1},\ldots,\gamma_{n_t,n_c}\right)$.
The $\alpha_{n,i}$ coefficients are computed as for (\ref{eq:closed_pairwise}).\\

\noindent
The asymptotic expression of  $P_w(\Delta,\Lambda)$ is
\begin{equation}\label{eq:asympt_BER_dec_bf}
P_w(\Delta,\Lambda)\eqv{N_{0}\rightarrow 0}
\C{2n_{r}n_tn_c-1}{n_rn_tn_c}
\prod_{k=1}^{n_c}\prod_{l=1}^{n_t}\left(\frac{2N_0}{\gamma_{k,l}^2}\right)^{n_r}
\end{equation}
The diversity associated with the considered pairs of Hamming weight $w$ is then
equal to the exponent $n_tn_cn_r$. The coding gain is given by the geometrical mean of the $\gamma_{k,l}^2$
and is equal to
\begin{equation}\label{eq:Gbf}
\mathcal{G}_{bf}(\Delta,\Lambda)=\left(\prod_{k=1}^{n_c}\prod_{l=1}^{n_t}\sum_{i=1}^{\kappa_{k,l}}d_{k,l,i}^{2}\right)^{1/(n_tn_c)}
\end{equation}
We will see in the following how to use this coding gain as a design criterion for the ST-BICM optimization.
We now consider an equivalent computation of the coding gain for a linearly precoded ST-BICM.

\subsection{Exact pairwise error probability for MIMO block fading channels with precoding}

When a linear precoder $\mat{S}$ of size $N_t\times N_t$ is used,
the detector computes soft outputs on the $N_t$ transmitted symbols
using the equivalent channel matrix $\mat{S}\mat{H}_k$ of size $N_t\times N_r$.
The structure of $\mat{H}_k$ is described in (\ref{eq:Hk_general}).
$\mat{S}\mat{H}_k$ can be seen as a correlated MIMO channel \cite{Veeravalli2001}.
Under the ideal interleaving condition, we consider at most a single
erroneous bit per block of $s$ time periods in position $1\leq\ell\leq mN_t$
inside the binary mapping of the transmitted symbol $\mat{z}$, leading to symbol $\bar{\mat{z}}^{\ell}$.
For simplicity reasons, we assume that the error weight $w$ satisfies $w\geq N_tN_c$.
Moreover, we assume that the mapping is mono-dimensional: the BSKs are transmitted on a single selected input
of the matrix $\mat{S}\mat{H}_{k}$.
Let $\LLR_{k,l,i}$ be the $i$-th variable among $\kappa_{k,l}$, 
corresponding to the transmission of a BSK on the equivalent $1\times N_r$
channel $\mat{S}_l\mat{H}_{k}$, where $\mat{S}_l$ corresponds to the $l$-th row of $\mat{S}$.
We have $\sum_{k=1}^{N_c}\sum_{l=1}^{N_t}\kappa_{k,l}=w$.
Let $d_{k,l,i}$ denote the BSK distance associated with $\LLR_{k,l,i}$.
We can use the factorization $\LLR_{k,l}=\sum_{i=1}^{\kappa_{k,l}}\LLR_{k,l,i}$
of all the LLR variables filtered with $\mat{S}_l\mat{H}_k$:
\begin{equation}
\LLR_{k,l} \sim \mathcal{N}\left(\frac{R_{k,l}}{2N_0},
\frac{R_{k,l}}{N_0}\right)
\end{equation}
where $R_{k,l}=\|\mat{V}_{k,l}\mat{H}_{k}\|^2$, $\mat{V}_{k,l}=\gamma_{k,l}\mat{S}_l$
and $\gamma_{k,l}^2=\sum_{i=1}^{\kappa_{k,l}}d_{k,l,i}^2$.
The variable $R_{k,l}$ is a generalized chi-square random variable with $2N_r$
correlated centered Gaussian components.
The random variable $\LLR_{k}=\sum_{l=1}^{N_t}\LLR_{k,l}$  satisfies
\begin{equation}
\LLR_{k} \sim \mathcal{N}\left(\frac{\sum_{l=1}^{N_t}R_{k,l}}{2N_0},\frac{\sum_{l=1}^{N_t}R_{k,l}}{N_0}\right)
\end{equation}
From appendix \ref{annex:A}, we get the following characteristic function:
\begin{eqnarray}
E_{\mat{H}_k}\left[\Psi_{\LLR_{k}}(j\nu)\right]=\prod_{t=1}^{n_s}\prod_{u=1}^{n_t}\left(1-\frac{\nu(j-\nu)}{2N_0}\vartheta_{k,u}^{[t]}\right)^{-n_r}
\end{eqnarray}
where $\vartheta_{k,u}^{[t]}$ is the $u$-th eigenvalue of
\begin{equation} \label{eq:sigma_prec}
\matgr{\Sigma}_k^{[t]}
=\sum_{l=1}^{N_t}\gamma_{k,l}^2{\mat{S}'}_l^{[t]*}{\mat{S}'}_l^{[t]}=
\sum_{l=1}^{N_t}\gamma_{k,l}^2\sum_{i=1}^{s/n_s}{\mat{S}}_{l}^{[t][i]*}{\mat{S}}_{l}^{[t][i]}
=\mat{M}_k^{[t]*}\mat{M}_k^{[t]}
\end{equation}
$\mat{M}_k^{[t]}$ is as an $n_t\times n_t$ Hermitian square root matrix of
$\matgr{\Sigma}_k^{[t]}$ and row vectors $\mat{S}^{[t][i]}_l$ of size $n_t$ and $s/n_s \times n_t$ matrices ${\mat{S}'}_l^{[t]}$
are defined from $\mat{S}$ as follows:
\begin{eqnarray}
\mat{S}=
\begin{array}{cccc}
~~\mat{S}_l^{[1]}(1\times N_t/n_s)&
&&
\mat{S}_l^{[n_s]}\\
\left[
\overbrace{
\begin{array}{ccc}
\mat{S}^{[1][1]}_1&\cdots&\mat{S}^{[1][s/n_s]}_1\\
\mat{S}^{[1][1]}_2&\cdots&\mat{S}^{[1][s/n_s]}_2\\
\vdots&&\\
\underbrace{\mat{S}^{[1][1]}_{N_t}}&\cdots&\underbrace{\mat{S}^{[1][s/n_s]}_{N_t}}
\end{array}}
\right.&
&
\begin{array}{c}
\cdots\\
\\
\\
\cdots
\end{array}
&
\left.
\overbrace{
\begin{array}{ccc}
\mat{S}^{[n_s][1]}_1&\cdots&\mat{S}^{[n_s][s/n_s]}_1\\
\mat{S}^{[n_s][1]}_2&\cdots&\mat{S}^{[n_s][s/n_s]}_2\\
\vdots&&\\
\underbrace{\mat{S}^{[n_s][1]}_{N_t}}&\cdots&\underbrace{\mat{S}^{[n_s][s/n_s]}_{N_t}}
\end{array}}
\right]
\end{array}\\
N_t/s=n_t\textrm{ coefficients} \nonumber
\end{eqnarray}
and
\begin{equation}
{\mat{S}'}_l^{[t]}=\left[
\begin{array}{c}
\mat{S}^{[t][1]}_l\\
\mat{S}^{[t][2]}_l\\
\vdots\\
\mat{S}^{[t][s/n_s]}_l\\
\end{array}
\right]
\end{equation}
The set of eigenvalues $\vartheta^{[t]}_{k,u}$ is a function of the precoding matrix $\mat{S}$ and the BSK distances set
$D^w$. 
Thanks to the independence of channel realizations for different $k$ values, we can multiply the characteristic functions:
\begin{eqnarray}
\Psi(j\nu)=\prod_{k=1}^{N_c}\prod_{t=1}^{n_s}\prod_{u=1}^{n_t}
\left(1-\frac{\nu(j-\nu)}{2N_0}\vartheta_{k,u}^{[t]}\right)^{-n_r}
\end{eqnarray}

Denote $\Delta=\{\delta_v\}$ the set of $n_\delta$ square-roots of non-null eigenvalues extracted from the sequence defined by
the $\vartheta_{k,u}^{[t]}$ values.
Each eigenvalue $\delta_v^2$ is repeated $\lambda_v$ times.
Observe that $n_\delta\leq n_c n_t$.
Finally, using the partial fraction expansion of $\Psi(j\nu)$ as for (\ref{eq:closed_pairwise}),
we obtain the exact pairwise error probability $P_w(\Delta,\Lambda)$ conditioned on $d_H(c,c')=w$:
\begin{equation}
P_{w}(\Delta,\Lambda)=\prod_{v=1}^{n_\delta}\left(-\frac{2N_0}{\delta_v^2}\right)^{\lambda_vn_r}
\sum_{v=1}^{n_\delta}\sum_{i=1}^{n_r\lambda_v}\frac{\alpha_{v,i}}{\left(\frac{1}{2}+\frac{1}{2}\sqrt{1+\frac{8N_0}{\delta_v^2}}\right)^i}
\end{equation}

\noindent
The asymptotic expression of  $P_w(\Delta,\Lambda)$ is
\begin{equation}\label{asympt_BER_dec_bf_linprec}
P_w(\Delta,\Lambda)\eqv{N_{0}\rightarrow 0}
\C{2n_{r}N_\delta-1}{n_rN_\delta}
\prod_{v=1}^{n_\delta}\left(\frac{2N_0}{\delta_v^2}\right)^{\lambda_vn_r}
\end{equation}
where $N_\delta = \sum_{v=1}^{n_\delta}\lambda_v$ is the total number of non-null eigenvalues.

The diversity associated with the considered pairs of Hamming weight $w$ is the
exponent equal to $\sum_{v=1}^{n_\delta}\lambda_v n_r = N_\delta n_r$. The coding gain is given by
\begin{equation}\label{Gs}
\mathcal{G}_{s,n_s}(\Delta,\Lambda)=\left(\prod_{v=1}^{n_\delta}\delta_v^{2\lambda_v}\right)^{1/N_\delta}
\end{equation}

We have derived for any signal-to-noise ratio an exact expression of the pairwise error probabilities
of a BICM with linear precoding, which is useful for a tight BER and FER estimation.
The asymptotic expression
leads to the well-known rank and determinant criteria \cite{Tarokh1998}\cite{Elgamal2003-2} for space-time code
optimization over MIMO block-fading channels, where the considered space-time code is the whole BICM structure.
As a remark, the asymptotic design criterion is usually derived by first
upperbounding the $Q(x)$ function by $\exp(-x^2/2)/2$ and then averaging
over the channel realizations. The obtained asymptotic expression has
a multiplying coefficient different from $\C{2n_{r}N_\delta-1}{n_rN_\delta}$, which
is inexact but provides the same design criterion.\\

Moreover, we notice that applying the Tarokh criterion  \cite{Tarokh1998} on the rank and determinant
to the precoder alone does not lead to the whole BICM optimization.
Quasi-optimal linear precoders will be designed to achieve full diversity and approach optimal
coding gain in section \ref{sec:linear-prec}.

\subsection{Evaluation of the Frame Error Rate}\label{subsec:evalfer}

For ergodic channels, the frame error rate is easily computed via a union bound. 
Indeed, only error events with minimum Hamming distance impact the error rate for a high signal-to-noise ratio 
and the observed diversity is equal to $n_rd_{Hmin}$.
For block-fading channels, the frame error rate computation is much more tricky since each pairwise error probability 
is supposed to have the full-diversity order $n_cn_tn_r$.
Due to the random nature of each eigenvalue in (\ref{Gs}),
it is difficult to know the impact of each distance configuration on the final FER.\\

However, one may assume that for a sufficiently high signal-to-noise ratio, the FER satisfies the following expression:
\begin{equation}
FER\simeq\sum_w A_w E_{(\Delta,\Lambda\mid w)}\left[P_w(\Delta,\Lambda)\right]
\end{equation} 
where  $A_w$ is weighting the impact of pairwise error probabilities
with Hamming weight $w$ in the global error probability
and the expectation on $(\Delta,\Lambda)$ is allowed by the interleaver random structure.
Let us define $\mathcal{G}$ the global coding gain. Since each pairwise error probability is supposed to
have full diversity, we write
\begin{equation}
FER\simeq \C{2n_rn_tn_c-1}{n_rn_tn_c}\left(\frac{2N_0}{ \mathcal{G}}\right)^{n_rn_tn_c}
\end{equation} 
and
\begin{equation}\label{eq:G}
 \mathcal{G} ^{-n_rn_tn_c}= \sum_w A_w E_{(\Delta,\Lambda\mid w)}\left[ \mathcal{G}(\Delta,\Lambda) ^{-n_rn_tn_c}\right]
\end{equation} 
where $ \mathcal{G}(\Delta,\Lambda) $ is the coding gain associated with one pair of codewords.
We note that optimizing independently all pairwise error probabilities,
which will be done in the following, enhances the global performance.
Moreover, we observe that the number of receive antennas does not affect
the coding gain of a single pairwise error probability. The effect of the receive diversity appears in
the expression of the global coding gain (\ref{eq:G}).
As $n_rn_tn_c$ grows, the smallest coding gains have more impact on the final performance. Asymptotically,
if $n_rn_tn_c\rightarrow +\infty$, only the nearest neighbors in the Euclidean code have an influence
on the FER, as for AWGN channels.\\

We will see in section \ref{sec:bicm_ideal}
that the best coding gain is achieved when all eigenvalues $\vartheta_{k,u}^{[t]}$ are equal.
In this ideal configuration, the coding gain is shown to be 
the same as with the same coded modulation
transmitted on a $1\times n_cn_tn_r$ quasi-static SIMO channel.
Simulating this latter case is less complex: the performance curve is semi-analytically computed
using a reference curve on an AWGN channel.
Alternatively, performance may be obtained by computing the Tangential Sphere Bound for spherical modulations \cite{Herzberg1996}.
In the following, \emph{ideal BICM} will refer to the performance of the ideal configuration,
which will be drawn on simulation results.
This lower bound has the advantage to take the modulation and error correcting code into account
and will be useful to evaluate the optimality of both the linear precoder and the channel interleaver.

\section{The Singleton bound with linear precoder \label{sec:singleton_bound}}

Definition \ref{def:ideal_interleaving} ensures that any pair
of codewords benefits from a full diversity order. In this section, we
derive a condition on the existence of a practical interleaver
that could achieve the conditions of Definition \ref{def:ideal_interleaving}.
Let us first make the following assumption:
\begin{hyp}\label{hyp:mimoconv}
The detector perfectly converts the $N_t\times N_r$ correlated MIMO $N_c$-block-fading channel $\mat{S}\mat{H}_k$
with QAM input into a $1\times sn_r$ SIMO $n_tn_c/s$-block-fading channel with BSK input, assuming that $s$
is a divisor of $n_tn_c$.
\end{hyp}
We will present in section \ref{sec:linear-prec} linear precoders that satisfy Assumption \ref{hyp:mimoconv}.
Under this condition, the detector collects an amount of diversity equal to $sn_r$.
The full diversity $n_t n_c n_r$ is collected by the detector when $s=n_t n_c$,
but unfortunately, the APP signal detection has an exponential complexity in $s$.
On the other hand, the BICM channel decoder is also capable of collecting a large
amount of diversity, but the latter is still limited by the
Singleton bound \cite{Knopp1997}\cite{Knopp2000}\cite{Malkamaki1999}. Hence, the lowest
complexity solution that reaches full diversity is to draw advantage
of the whole channel code diversity
and recover the remaining diversity by linear precoding.
The best way to choose the spreading factor $s$ is given by the
Singleton bound described hereafter.\\

The studied ST-BICM is a serial concatenation of a rate $R_\mathcal{C}$ binary convolutional code
$\mathcal{C}$, an interleaver of size $N_\mathcal{C}L_\mathcal{C}$ bits,
and a QAM mapper followed by the precoder as described in section~\ref{sec:system}.
When $\mat{S}$ is the identity matrix, the ST-BICM diversity order is upper-bounded by \cite{Knopp2000}:
\begin{equation}\label{eqsb_without}
 \Upsilon \le n_r \left( \lfloor n_cn_t(1-R_c)\rfloor +1 \right)
\end{equation}
The maximal diversity given by the outage limit under a finite size QAM alphabet 
also achieves the above Singleton bound \cite{Guillen2004}.
With a vanishing coding rate, i.e., $R_c \leq 1/(n_cn_t)$, 
it is possible to attain the overall system diversity order $n_rn_cn_t$
produced by the receive antennas, the transmit antennas and the distinct channel states.
Unfortunately, this is unacceptable due to the vanishing transmitted information rate.
Precoding is one means to achieve maximum diversity with a non-vanishing coding rate.\\

The integer $N_b=n_c n_t/s$ is the best diversity multiplication factor to be collected by $\mathcal{C}$.
The length of a $\mathcal{C}$ codeword is $L_\mathcal{C} N_\mathcal{C}$ binary elements.
Let us group $L_\mathcal{C} N_\mathcal{C}/N_b$ bits into one non-binary symbol
creating a non-binary code $\mathcal{C}'$.
Now, $\mathcal{C}'$ is a length-$N_b$ code built on an alphabet
of size $2^{L_\mathcal{C} N_\mathcal{C}/N_b}$.
The Singleton bound on the minimum Hamming distance of the non-binary $\mathcal{C}'$
becomes $D_H \le N_b-\lceil N_b R_\mathcal{C} \rceil+1$.
Multiplying the previous inequality with the
Nakagami law order $s n_r$ yields the maximum achievable diversity order
after decoding \cite{Gresset2004-ISIT}:
\begin{equation}
\label{equ_singletonbounddiv}
\Upsilon \leq s n_r\left\lfloor\frac{n_c n_t}{s}(1-R_\mathcal{C})+1\right\rfloor
\end{equation}
Finally, since $\Upsilon$ is upper-bounded
by the channel intrinsic diversity and the minimum Hamming distance $d_{Hmin}$ of the binary code,
we can write
\small
\begin{equation}
\label{equ_upperbounddiv}
\Upsilon \leq \min \left( s n_r\left\lfloor\frac{n_c n_t}{s}(1-R_\mathcal{C})+1\right\rfloor;n_t n_c n_r;s n_r d_{Hmin} \right)=\Upsilon_{max}
\end{equation}
\normalsize

If $d_{Hmin}$ is not a limiting factor (we choose $\mathcal{C}$ accordingly),
we can select the value of $s$ that leads to a modified Singleton bound greater than or equal to $n_t n_c n_r$.

\begin{prop}\label{prop:singletonbound}
Considering a BICM with a rate $R_\mathcal{C}$ binary error-correcting code
on an $n_t\times n_r$ MIMO channel with $n_c$ distinct channel states per codeword,
the spreading factor $s$ of a linear precoder must
be a divisor of
$n_t n_c$ and must satisfy $s \ge R_\mathcal{C} n_c n_t$ in order to achieve the full
diversity $n_t n_c n_r$ for any pair of codewords.
In this case, the ideal interleaving conditions
can be achieved with an optimized interleaver.
\end{prop}

The smallest integer $s_{opt}$ satisfying the above proposition minimizes the detector's complexity.
If $R_\mathcal{C} >1/2 $, then $s_{opt}=n_c n_t$ which involves the highest complexity.
If $R_\mathcal{C} \le 1/(n_c n_t)$, linear precoding is not required.\\
Tables \ref{tab:table1} and \ref{tab:table2} show the diversity order
derived from the Singleton bound versus $s$ and $n_t$,
for $n_c=1$ and $n_c=2$ respectively.
The values in bold indicate full diversity
 configurations. For example, in Table \ref{tab:table1}, for $n_t=4$,
$s=2$ is a better choice than $s=4$ since it leads to an identical diversity order
with a lower complexity.

%
\section{Linear precoder optimization \label{sec:linear-prec}}

Many studies have been published on space-time
spreading matrices introducing some redundancy, well-known as space-time block codes.
On one hand, some of them  are decoded by a low-complexity ML decoder, but they
sacrifice transmission data rate for the sake of high performance.
Among them, the Alamouti scheme \cite{Alamouti1998} is the most famous, but is only optimal for a
$2\times 1$ MIMO channel. The other designs allowing for low ML decoding complexity are
based on an extension of the Alamouti principle (e.g., DSTTD \cite{TI2001-DSTTD}) but also sacrifice the data rate.
On the other hand, full rate space-time codes have recently been proposed
\cite{belfiore2004}\cite{Damen2002}\cite{Damen2003}\cite{Damen2003-2}\cite{Dayal2003}\cite{Elgamal2003}\cite{Oggier2004}.
However, their optimization does not take into account their concatenation with an error correcting code.
In this section, we describe a near-ideal solution for linear precoding in BICMs under iterative decoding process.
Our strategy is to separate the coding step and the geometry properties
in order to express some criteria allowing the construction
of a space-time spreading matrix for given channel parameters $n_t$, $n_r$ and $n_c$.
The inclusion of rotations to enhance the BICM performance over single antenna
channels has been proposed in \cite{Lamythese}. Our solution 
uses this concept for designing a space-time code including a powerful error correcting code.\\

When the channel is quasi-static or block-fading with parameter $n_c$, the diversity is upper bounded by
$n_cn_tn_r$ which may be more limiting than $n_rd_{Hmin}$ (e.g., $n_t=2$, $n_r=1$, $n_c$=1).
We introduce a new design criterion of space-time spreading matrices that guarantees a
diversity proportional to the spreading factor, within the upper-bound, and a maximal coding gain at the last iteration
of an iterative joint detection and decoding.

\subsection{Coding gain under both ideal interleaving and precoding\label{sec:bicm_ideal}}

First we look for the best achievable coding gain for the fixed parameters $n_t$, $n_r$, $n_c$, $R_\mathcal{C}$ and
the appropriate way to choose the error correcting code, the binary mapping, the linear precoder
and its parameters $s$ and $n_s$ to achieve the ideal coding gain.\\

We want to achieve full diversity under ML decoding or iterative joint detection and decoding,
this induces that there are $n_cn_t$ non-null eigenvalues $\vartheta_{k,u}^{[t]}$ (see (\ref{asympt_BER_dec_bf_linprec})):
\begin{equation}\label{eq:gs}
\mathcal{G}_{s,n_s}(\Delta,\Lambda)=\left(\prod_{k=1}^{N_c}\prod_{t=1}^{n_s}\prod_{u=1}^{n_t}
{\vartheta_{k,u}^{[t]}}\right)^{1/(n_cn_t)}
\end{equation}

Furthermore, we want to maximize the $\mathcal{G}_{s,n_s}(\Delta,\Lambda)$ expression.
Assuming that each row $\mat{S}_l$ is normalized to $1$, we get
\begin{equation}\label{eq:norm_cons}
\sum_{k=1}^{N_c}\sum_{t=1}^{n_s}\sum_{u=1}^{n_t}\vartheta_{k,u}^{[t]}
=\sum_{k=1}^{N_c}\sum_{l=1}^{N_t}\gamma_{k,l}^2
=\sum_{k=1}^{N_c}\sum_{l=1}^{N_t}\sum_{i=1}^{\kappa_{k,l}}d_{k,l,i}^2
\end{equation}
Under this condition, the ideal coding gain is achieved when all eigenvalues are equal
\begin{equation}\label{eq:trueprop}
\vartheta_{k,u}^{[t]}=\sum_{k'=1}^{N_c}\sum_{l=1}^{N_t}\frac{\gamma_{k',l}^2}{n_tn_{c}} \quad \forall (k,t,u)
\end{equation}
which leads to
\begin{equation}\label{eq:optgain}
\mathcal{G}_{ideal}(\Delta,\Lambda)
=\sum_{k=1}^{N_c}\sum_{l=1}^{N_t}\frac{\gamma_{k,l}^2}{n_tn_{c}}
=\sum_{i=1}^{w} \frac{d_i^2}{n_tn_{c}}
\end{equation}
The exact pairwise error probability expression simplifies to the
classical expression of the performance of a BPSK with
distance $\sum_{j=1}^{w}d_j^2$ over a diversity channel with order $n_cn_tn_r$ \cite{Proakis2001}:
\small
\begin{equation}\label{eq:idealpairwise}
P_{w, ideal}(\Delta,\Lambda)=\left(1-\left(1+\frac{8N_0n_tn_{c}}{\sum_{j=1}^{w}d_j^2}\right)^{-1/2}\right)^{n_cn_tn_r}
\sum_{k=0}^{n_cn_tn_r-1}\frac{\C{n_cn_tn_r+k-1}{k}}{2^{n_cn_tn_r+k}}       
\left(1+\left(1+\frac{8N_0n_tn_{c}}{\sum_{j=1}^{w}d_j^2}\right)^{-1/2}\right)^{k}
\end{equation}
\normalsize

As stated in the introduction, in an ST-BICM, precoded modulation symbols 
quantify the Shannon sphere
and best quantization is obtained by uniformly distribute them on the sphere.
After transmission on a fading channel,
vectors belong to an ellipsoid obtained by applying an homothety on the sphere.
From (\ref{eq:trueprop}) and (\ref{eq:optgain}), we see that the ideal coding gain
is obtained by equally distributing the Euclidean distance between two codewords 
among the $n_tn_c$ channel states. Hence, the Euclidean distance varies as a $n_tn_cn_r$ Nakagami distribution, 
according to the square norm of the ellipsoid axes.
Thus, an ideal ST-BICM aims at uniformly distributing the precoded modulation
symbols, whatever the channel realization, i.e., whatever the homothety.
The ideal coding gain is a fundamental limit which cannot be outperformed.
It is useful to evaluate how optimal the practical design of a BICM is.
We aim at finding the best design, corresponding to eigenvalues which are as close to each other
as possible.
The more different from each other the eigenvalues are, the lower the product in (\ref{eq:gs}) and the coding gain are.
From (\ref{eq:optgain}), we see that the ideal coding gain is 
the same as for the same coded modulation
transmitted on a $1\times n_cn_tn_r$ single-input multiple-output (SIMO) channel, 
applying the appropriate $E_b/N_0$ normalization.\\

Without linear precoding, the ideal coding gain
is only achieved if all $\gamma_{k,l}$ are equal.
Remember that each $\gamma_{k,l}$ is a sum of $\kappa_{k,l}$ distances $d_{k,l,i}$.
Thanks to the second point in Definition \ref {def:ideal_interleaving}, 
the $\kappa_{k,l}$ values are close to $w/(n_tn_c)$
and their variance decreases when $w$ increases.
Thus, with a powerful error correcting code having minimum Hamming distance much greater than
$n_tn_c$ and $|D|$, 
each $\gamma_{k,l}$ value is almost equal to the average 
$\sum_{k=1}^{N_c}\sum_{l=1}^{N_t}\sum_{i=1}^{\kappa_{k,l}}d_{k,l,i}^2/(N_tN_c)$ of $d_{k,l,i}$ values
and quasi-ideal coding gain is observed.\\

If the error correcting code is not powerful enough to achieve the ideal coding gain, i.e., the
$\gamma_{k,l}$ values are very different,
the linear precoder provides an additional coding gain by averaging the $\gamma_{k,l}$ values,
as we will see in the following.
First, we derive the optimal coding
gain which can be achieved using an ideal linear precoder for a given binary labeling and error correcting code.
Variables $\gamma_{k,l}$ for different $k$ values correspond to independent
channel realizations $\mat{H}_k$ which are not linked by the linear precoder.
Thus, random
variables $\prod_{t=1}^{n_s}\prod_{u=1}^{n_t}{\vartheta_{k,u}^{[t]}}$
are independent for distinct values of $k$. 
The optimal coding gain with linear precoding is 
\begin{equation}\label{eq:Gopt}
\mathcal{G}_{s,n_s,opt}(\Delta,\Lambda)=\prod_{k=1}^{N_c}
\left(\sum_{l=1}^{N_t}\frac{\gamma_{k,l}^2}{n_tn_s}\right)^{1/N_c}
\end{equation}
Equation (\ref{eq:Gopt}) means that an optimal linear precoder is capable
of making eigenvalues equal for a same $k$.
However, for different values of $k$, eigenvalues $\vartheta_{k,u}^{[t]}$ 
are different, which induces a coding gain loss.
When the mapping and error correcting code are given and the interleaving is ideal,
the choice of linear precoding parameters impacts on optimal coding gain.   
Let us consider codewords that are equidistant from the transmitted codeword, i.e.,
a set of distance configurations corresponding to a same value of $\sum_{k,l}\gamma_{k,l}^2$.
The variance of $\sum_{l=1}^{N_t}\gamma_{k,l}^2/(n_tn_s)$ over this set
decreases when $n_s$ increases, as the number of distances building each $\gamma_{k,l}$ is higher.
The lower the variance of eigenvalues, the higher the coding gain.
Thus, $\mathcal{G}_{s,n_s,opt}(\Delta,\Lambda)$ is an increasing function of $n_s$
and, for a given $s$, we should choose $n_s=\min(s,n_c)$.
The optimal coding
gain $\mathcal{G}_{s,\min(s,n_c),opt}(\Delta,\Lambda)$ is an increasing function of $s$.
If $n_s=n_c$, which implies $s=n_tn_c$, the ideal coding gain is achieved by the optimal precoder.
Finally, we can surround the coding gain at full diversity as follows: 
\begin{equation}
\forall s \quad \mathcal{G}_{ideal}(\Delta,\Lambda)
\geq\mathcal{G}_{s,\min(s,n_c),opt}(\Delta,\Lambda)
\geq\mathcal{G}_{s,1,opt}(\Delta,\Lambda)
\geq\mathcal{G}_{1,1,opt}(\Delta,\Lambda)
\geq\mathcal{G}_{bf}(\Delta,\Lambda)
\end{equation}

If, for any pairwise error probability, $\mathcal{G}_{bf}(\Delta,\Lambda)\simeq\mathcal{G}_{ideal}(\Delta,\Lambda)$,
the linear precoder optimization is useless from a coding gain point-of-view.
However, obtaining near-ideal coding gain without precoding requires 
an optimization of the error correcting code and mapping
for any pairwise error probability, which is intractable for non-trivial modulations
and codes.
Furthermore, the first objective of linear precoding is the diversity control, which
has a high influence on the performance even at medium FER ($10^{-2}\sim10^{-3}$),
especially for low diversity orders. 
Therefore, precoding is often useful in the BICM structure.\\

After the impact of the linear precoding for a given pairwise error
probability, let us consider the behavior of the global performance under linear precoding.
As stated in section \ref{subsec:evalfer}, if $n_rn_tn_c$ grows, the pairs of codewords 
providing the smallest coding gains have more impact on the final performance. 
Since the linear precoder provides a more
substantial gain for the low Hamming weight configurations,
the coding gain of the linear precoder will be magnified as the diversity grows.

\begin{figure}[!ht]
   \centering
   \includegraphics[width=0.7\columnwidth]{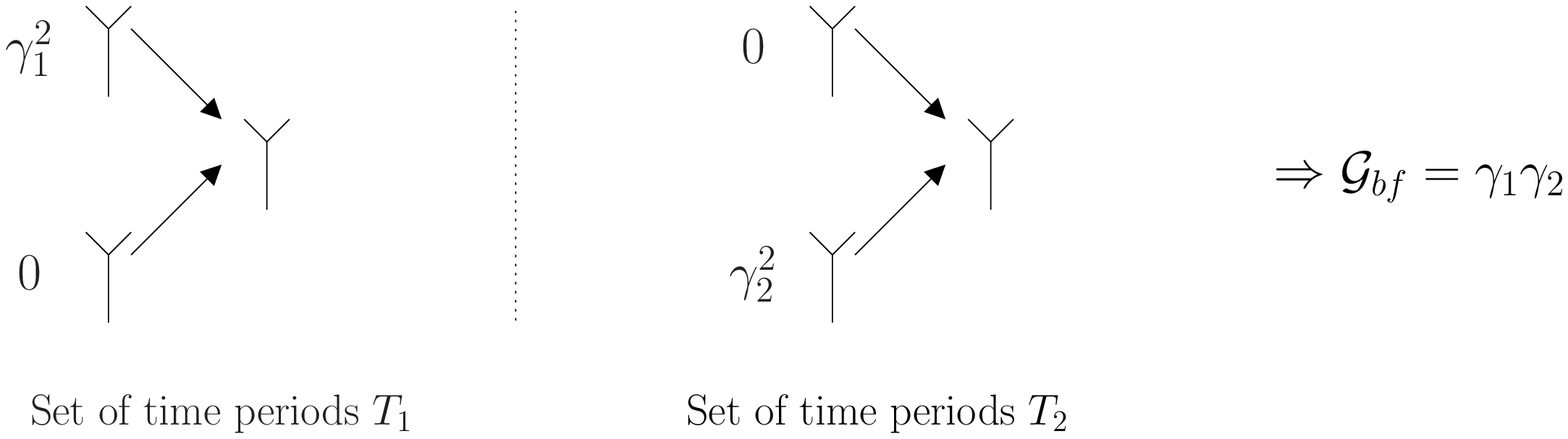}
   \caption{Coding gain for unprecoded $2\times 1$ quasi-static MIMO channel}
   \label{fig:example_pre}
\end{figure}

\begin{figure}[!ht]
   \centering
   \includegraphics[width=0.7\columnwidth]{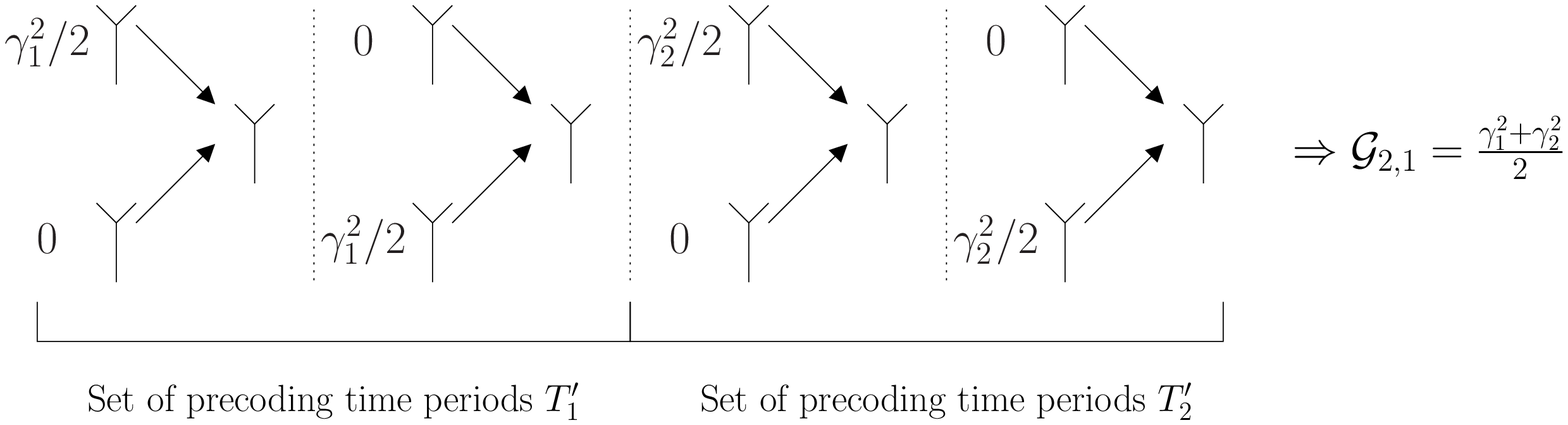}
   \caption{Coding gain for precoded $2\times 1$ quasi-static MIMO channel, $s=2$}
   \label{fig:example_pre2}
\end{figure}

\paragraph{Example of ideal coding gain:}
In order to illustrate the role of the linear precoding in the coding gain optimization, we consider a
$2\times 1$ quasi-static MIMO channel and a pairwise error probability between two codewords separated by
a Hamming distance of $w$ bits.
Fig. \ref{fig:example_pre} represents the distribution of the two $\gamma_1$ and $\gamma_2$ values over
the two transmit antennas without linear precoding. 
Bits transmitted on antennas 1 and 2 are transmitted on the sets of time periods $T_1$ and $T_2$, respectively.
Thanks to ideal interleaving, $T_1 \bigcap T_2 = \emptyset$.
This illustrates
the factorization of the distances into the $\gamma$ values. The instantaneous coding gain is equal to $\sqrt{\gamma_1^2\gamma_2^2}$.
Now let us consider a specific linear precoder, which spreads the values $\gamma_1$ and $\gamma_2$ 
as presented in Fig. \ref {fig:example_pre2} 
over two time periods and two transmit antennas dividing the squared distance
in two equal parts $\gamma_1^2/2$ and $\gamma_2^2/2$ respectively.
The average value $(\gamma_1^2+\gamma_2^2)/2$ is transmitted on each antenna, the coding gain
is optimal and equal to $(\gamma_1^2+\gamma_2^2)/2$.
For example, consider a BPSK modulation and a pairwise error probability
with Hamming weight $3$. With optimal linear precoding,
the ideal interleaving provides for example $\gamma_1^2=2\times 2^2$
and $\gamma_2^2=1\times 2^2$. With optimal linear precoding, we have
a distance $(2\times2^2+1\times 2^2)/2$ associated with each antenna.
The ratio between the two averaged coding gains is equal to $\sqrt{9/8}$,
i.e., we expect a gain of $0.26$~dB when using linear precoding.
With $w=5$ and $w=11$, the coding gain becomes $10\log_{10}(\sqrt{24/25})\simeq 0.09$~dB
and $10\log_{10}(\sqrt{120/121})\simeq 0.02$~dB, respectively.
The higher the Hamming weight involved in the pairwise error probability is,
the less the coding gain provided by linear precoding is. $\blacksquare$\\

We see on Table \ref{tab:gain} the best gain to be provided by linear precoding for a quasi-static channel with
BPSK input with respect to a full diversity unprecoded scheme. 
These gains are particularly low because the error correcting code aims at recovering a large amount of coding gain.
This illustrates that BICMs are very efficient transmission schemes.
As a remark, if a modulation with higher spectral efficiency is used with Gray mapping,
the nearest neighbor in the Euclidean code has the same distance configuration as if a BPSK modulation was used.
Moreover, for high diversity orders, the global error rate for high $E_b/N_0$ will be dominated by the neighbors and
the gain provided by linear precoding will be very close to the ones shown in Table \ref{tab:gain}.
However, if the diversity is low, 
the gains provided by linear
precoding may be much more important. Assume that a 16-QAM modulation with Gray mapping is transmitted
on a $n_t=2$ quasi-static channel. For instance, if $w=5$, there exists a neighbor with distance configuration $(3A,3A,3A,A,A)$
(e.g., see \cite{Gresset2004-IT}),
and $\gamma_1^2=9A^2+9A^2+9A^2$, $\gamma_2^2=A^2+A^2$.
The gain to be provided by linear precoding is equal to $10\log_{10}(29/2/\sqrt{54})=2.95$~dB.
As already stated, the final coding gain is equal to a weighted sum
of all the coding gains, where the weighting coefficients cannot be easily computed in the case of low diversity orders.\\

Even if linear precoding does not always provide a substantial coding gain,
its prior aim is the diversity order control.
Thus, we will focus on the design of linear precoders aiming at reaching full diversity
and maximizing the coding gain for any set of parameter $(n_t,s,n_s)$.

\subsection{A new class of linear precoders}\label{sec:class_lp}

Under linear precoding, the optimal coding gain is achieved if all
$\vartheta_{k,u}^{[t]}$ variables are equal for a same $k$.
Let us first consider the eigenvalues associated with the independent
realizations in the spreading matrix, indexed by $t$.
First, two matrices $\mat{M}_k^{[t_1]}$ and $\mat{M}_k^{[t_2]}$, as introduced in (\ref{eq:sigma_prec}), should have the
same eigenvalues, which is satisfied if
$\forall (t_1,t_2), \mat{M}_k^{[t_1]}=\mat{R}^{t_1,t_2*}\mat{M}_k^{[t_2]}\mat{R}^{t_1,t_2}$, where $\mat{R}^{t_1,t_2}$
is a unitary matrix, for example a rotation.
Hence,
$\forall (t_1,t_2), {\mat{S}'}_l^{[t_1]}={\mat{S}'}_l^{[t_2]}\mat{R}^{t_1,t_2}$.
The precoding sub-part ${\mat{S}}_l^{[t_1]}$, with spreading factor $s'=s/n_s$, experiences a quasi-static channel.
We assume that $s'$ is an integer, divisor of $n_t$.
It is sufficient to design the first sub-part of the precoder matrix rows for a quasi-static channel 
and rotate it to compute the other sub-parts.
Furthermore, any choice of $\mat{R}^{t_1,t_2}$ leads to the same performance because the eigenvalues remain unchanged.
The condition simplifies to $\|{\mat{S}'}_l^{[t_1]}\|=\|{\mat{S}'}_l^{[t_2]}\|$.\\

Let us now optimize for a given index $t$ the equivalent
precoder over the quasi-static channel $\diag\left(\mat{H}_k^{[t]},\ldots,\mat{H}_k^{[t]}\right)$,
in which $\mat{H}_k^{[t]}$ is repeated $s'$ times.
If all the eigenvalues of $\mat{M}_k^{[t]}\mat{M}_k^{[t]*}$ are equal,
$\mat{M}_k^{[t]}$ and $\mat{M}_k^{[t]*}$ are weighted unitary matrices and
\begin{equation}
\mat{M}_k^{[t]}\mat{M}_k^{[t]*}=\mat{M}_k^{[t]*}\mat{M}_k^{[t]}=\sum_{l=1}^{N_t} \gamma_{k,l}^2\sum_{i=1}^{s'}{\mat{S}}_{l}^{[t][i]*}{\mat{S}}_{l}^{[t][i]}
\end{equation}
Matrix ${\mat{S}}_{l}^{[t][i]*}{\mat{S}}_{l}^{[t][i]}$ has rank one and matrix
$\sum_{i=1}^{s'}{\mat{S}}_{l}^{[t][i]*}{\mat{S}}_{l}^{[t][i]}$ has maximum rank $s'$. 
If $s'<n_t$, it can be shown
that it is impossible to get all eigenvalues equal to $\sum_{l=1}^{N_t} \gamma_{k,l}^2/{n_tn_s}$ as required to achieve the
optimal coding gain. However, in order to insure that $\mat{M}_k^{[t]}\mat{M}_k^{[t]*}$ has a rank $n_t$
and that the eigenvalues are as equal as possible,
we group $ss'$ values $\gamma_{k,l}$ together and associate them with one of the $n_t/s'$ groups of
$s'$ eigenvalues:
we denote ${\mat{S}}_{l}^{[t][i][j]}$ the $j$-th sub-part of size $s'$ of  ${\mat{S}}_{l}^{[t][i]}$
and $\{l_2,l_1\}$ the index of the $(l_2-1)ss'+l_1$-th row of $\mat{S}$,
where $l_2\in[1,n_t/s'], l_1\in[1,ss']$.
Let us assume that ${\mat{S}}_{\{l_2,l_1\}}^{[t][i]}$ has only one non-null sub-part in position $l_2$, i.e.,
\begin{equation}
\forall j\neq l_2
\quad \mat{S}_{\{l_2,l_1\}}^{[t][i][j]}=[0,\ldots,0]
\end{equation}
Considering such a structure is equivalent to considering separate precoding
on $n_t/s'$ distinct groups of $s'$ transmit antennas.
We have
\begin{eqnarray}
\sum_{l=1}^{N_t}\gamma_{k,l}^2 {\mat{S}'}_{l}^{[t]*}{\mat{S}'}_{l}^{[t]}=
\sum_{l_2=1}^{n_t/s'}\sum_{l_1=1}^{ss'}\gamma_{k,\{l_2,l_1\}}^2
\sum_{i=1}^{s'}{\mat{S}}_{\{l_2,l_1\}}^{[t][i]*}{\mat{S}}_{\{l_2,l_1\}}^{[t][i]}\\=
\sum_{l_2=1}^{n_t/s'}\sum_{l_1=1}^{ss'}\gamma_{k,\{l_2,l_1\}}^2
\mathfrak{D}_{l_2}\left(\sum_{i=1}^{s'}{\mat{S}}_{\{l_2,l_1\}}^{[t][i][l_2]*}{\mat{S}}_{\{l_2,l_1\}}^{[t][i][l_2]}\right)
\end{eqnarray}
where $\mathfrak{D}_{l_2}(A)$ is a block diagonal matrix with only one non-null block $A$ in position $l_2$.
We choose
${\mat{S}}_{\{l_2,l_1\}}^{[t][i][l_2]}$ proportional to the $i$-th row of a $s'\times s'$ unitary matrix,
such as $\|{\mat{S}}_{\{l_2,l_1\}}^{[t][i][l_2]}\|^2=1/s$:
\begin{eqnarray}
\sum_{l=1}^{N_t}\gamma_{k,l}^2 {\mat{S}'}_{l}^{[t]*}{\mat{S}'}_{l}^{[t]}&=&
\sum_{l_2=1}^{n_t/s'}\sum_{l_1=1}^{ss'}\gamma_{k,\{l_2,l_1\}}^2
\mathfrak{D}_{l_2}\left(\frac{1}{s}\mat{I}_{s'}\right)\\
&=&\frac{1}{s}
\sum_{l_1=1}^{ss'}\diag\left(\gamma_{k,\{1,l_1\}}^2\mat{I}_{s'},\ldots,
\gamma_{k,\{n_t/s',l_1\}}^2\mat{I}_{s'}\right)
\end{eqnarray}
which leads to
\begin{equation}
l_2\leq n_t/s',u\leq s',\quad \vartheta_{k,(l_2-1)ss'+u}^{[t]}=
\frac{1}{s}\sum_{l_1=1}^{ss'}\gamma_{k,\{l_2,l_1\}}^2
\end{equation}
The random variables $\gamma_{k,\{l_2,l_1\}}^2$ are independent and identically distributed for different values of
$l_1$ and $l_2$, the coding gain is 
\begin{equation}\label{eq:Glinprec}
\mathcal{G}_{s,n_s}(\Delta,\Lambda)=
\prod_{k=1}^{N_c}\prod_{l_2=1}^{n_t/s'}
\left(\sum_{l_1=1}^{ss'}\frac{\gamma_{k,\{l_2,l_1\}}^2}{s}\right)^{s'/(N_cn_t)}
\end{equation}

For any value of $n_s$, the gain expressed in (\ref{eq:Glinprec}) is a geometric mean of
order $n_tn_c/s$.
For a given realization $\{d_1,\ldots,d_w\}$, a given $s$ and for any $n_s$,
$\sum_{k=1}^{N_c}\sum_{l_2=1}^{n_t/s'}\sum_{l_1=1}^{ss'}\gamma_{k,\{l_2,l_1\}}^2$
and thus $\sum_{l_1=1}^{ss'}\gamma_{k,\{l_2,l_1\}}^2$ are constant, 
ensuring the same coding gain.
However, such a precoder does not achieve the optimal coding gain for any value of $s'$.
The summation is made over $ss'$ different values
whereas the optimal coding gain in (\ref{eq:Gopt}) necessitates a summation over $sn_t$ values.
Only if $ss'$ is high enough, the obtained coding gain
is almost optimal.
If $s'=n_t$, the complete spatial transmit diversity is collected by the detector and the optimal
coding gain is achieved.

\begin{prop}\label{prop:class_prec}
\textbf{Dispersive Nucleo Algebraic (DNA) Precoder}
Let $\mat{S}$ be the $N_t\times N_t$ precoding matrix of a BICM over a $n_t\times n_r$ MIMO $n_c$-block-fading channel.
Assume that $\mat{S}$ precodes a channel block diagonal matrix with $s$ blocks and $n_s$ channel realizations.
We denote $s$ the spreading factor, $N_t=sn_t$ and $s'=s/n_s$. Let $\mat{S}_l^{[t]}$ be the $t$-th sub-part of size $N_t/n_s$
of the $l$-th row of $\mat{S}$. Let $\mat{S}_l^{[t][i]}$ be the $i$-th sub-part of size $n_t$ of $\mat{S}_l^{[t]}$.
Let $\mat{S}_l^{[t][i][j]}$ be the $j$-th sub-part of size $s'$ of $\mat{S}_l^{[t][i]}$.
The sub-part $\mat{S}_l^{[t][i][j]}$ is called nucleotide.
The linear precoder guarantees full diversity
and quasi-optimal coding gain at the decoder output under maximum likelihood decoding of the BICM if it satisfies the two conditions
of null nucleotides and orthogonal nucleotides for all $t\in[1,n_s], i\in[1,s'], l_1 \in[1,ss'], l_2\in[1,n_t/s']$
and $\{l_2,l_1\}=(l_2-1)ss'+l_1$:
\begin{equation}
\left\{
\begin{array}{llr}
\forall j\neq l_2,j\in[1,n_t/s'],& \mat{S}_{\{l_2,l_1\}}^{[t][i][j]}=0_{1\times s'}&\textrm{\underline{Null Nucleotide condition}}\\
~\\
\forall i'\neq i, i'\in[1,s'], &{\mat{S}}_{\{l_2,l_1\}}^{[t][i][l_2]}{\mat{S}}_{\{l_2,l_1\}}^{[t][i'][l_2]*}
=\frac{1}{s}\mathfrak{d}(i-i')&\textrm{\underline{Orthogonal Nucleotide condition}}
\end{array}
\right.
\end{equation}
\end{prop}
where $\mathfrak{d}(0)=1$ and  $\mathfrak{d}(x\neq0)=0$.\\

Let us take for example $n_t=4$, $n_s=1$ and $s=2$. A DNA matrix would have the following structure:
\begin{equation}
\mathrm{DNA}(n_t=4,n_s=1,s=2)=\left[
\begin{array}{cccc}
\mat{S}_{\{1,1\}}^{[1][1][1]}&0&\mat{S}_{\{1,1\}}^{[1][2][1]}&0\\
\mat{S}_{\{1,2\}}^{[1][1][1]}&0&\mat{S}_{\{1,2\}}^{[1][2][1]}&0\\
\mat{S}_{\{1,3\}}^{[1][1][1]}&0&\mat{S}_{\{1,3\}}^{[1][2][1]}&0\\
\mat{S}_{\{1,4\}}^{[1][1][1]}&0&\mat{S}_{\{1,4\}}^{[1][2][1]}&0\\
0&\mat{S}_{\{2,1\}}^{[1][1][2]}&0&\mat{S}_{\{2,1\}}^{[1][2][2]}\\
0&\mat{S}_{\{2,2\}}^{[1][1][2]}&0&\mat{S}_{\{2,2\}}^{[1][2][2]}\\
0&\mat{S}_{\{2,3\}}^{[1][1][2]}&0&\mat{S}_{\{2,3\}}^{[1][2][2]}\\
0&\mat{S}_{\{2,4\}}^{[1][1][2]}&0&\mat{S}_{\{2,4\}}^{[1][2][2]}
\end{array}
\right]
\end{equation}

Now, let us consider a linear precoder matrix $\mat{S}$ that satisfies Proposition \ref{prop:class_prec}.
We build a $ss'\times N_r$ matrix $\mat{H}_k^{\{i\}}$ from the rows of $\mat{H}_k$ corresponding
to the $i$-th group of $s'$ transmit antennas.
Elements of $\mat{H}_k^{\{i\}}$ are defined as follows:
\begin{eqnarray}
\forall i = 0 \ldots n_t/s'-1, \forall j = 0 \ldots s'-1, \forall u = 0 \ldots s-1, \forall v = 0 \ldots N_r-1, \nonumber \\
\mat{H}_k^{\{i\}}(j+us',v)=\mat{H}_k(j+un_t+is',v)
\end{eqnarray}
Likewise, $\mat{S}^{\{i\}}$ is the $ss'\times ss'$ matrix obtained from the $i$-th block of $ss'$ rows of $\mat{S}$ and every $n_t/s'$-th
block of $s'$ columns beginning with the $i$-th block.
We easily show that
\begin{equation}
\mat{S}\mat{H}_k=
\left[
\begin{array}{ccc}
\mat{S}^{\{1\}}\mat{H}_k^{\{1\}}\\
\mat{S}^{\{2\}}\mat{H}_k^{\{2\}}\\
\vdots\\
\mat{S}^{\{n_t/s'\}}\mat{H}_k^{\{n_t/s'\}}\\
\end{array}
\right]
\end{equation}
which means that the matrix $\mat{S}$ independently precodes the $n_t/s'$  groups of transmit antennas.

Thus, the optimization may be split into $n_t/s'$ independent optimizations of linear precoders for $s' \times n_r$ MIMO
$n_s$-block-fading channels with linear spreading factor $s$. As $s=s'n_s$, full space-time spreading 
of the $s' \times n_r$ block-fading channel is performed, i.e., the maximum diversity order $sn_r$ is
collected by the detector.\\

From (\ref{eq:Gbf}) and (\ref{eq:Glinprec}), we notice that, at the decoder input and under ideal interleaving condition,
the linear precoder at the transmitter end and the detector at the receiver end
allow the conversion of the $n_t\times n_r$ MIMO channel with $n_c$ independent blocks into a
$1\times sn_r$ SIMO channel with $n_cn_t/s$ independent blocks with BSK input.
The independence of the blocks is provided by the structure of the linear precoding matrix:
\begin{enumerate}
\item The null nucleotides dispatch the transmitted symbols
on $n_t/s'$ different blocks of $s'$ antennas.
\item The orthogonal nucleotides provide full diversity
and a coding gain increasing with the spreading factor.
\end{enumerate}

For instance, if a rate 1/2 BICM is transmitted on a quasi-static $4\times 2$ MIMO channel, 
linear precoding with $s=2$ is required to achieve full diversity:
a full-rate space-time block code with spreading factor $s=s'=2$ may independently be applied on 2 separate groups of 2 transmit antennas.
Good $2\times2$ space-time block codes are for instance the TAST \cite{Damen2003} 
and the Golden code \cite{belfiore2004}.\\

Assume that $n_c=1$, $n_t=n_r=2$ and $s=s'=2$.
The Golden code is the best space-time code for uncoded $2\times 2$ quasi-static MIMO channels.
However, it does not satisfy the equal norm property of orthogonal nucleotides in Proposition~\ref{prop:class_prec}.
Indeed, one row of the Golden linear precoder contains two non-null coefficients of square norm $\alpha_1=0.277$ and
$\alpha_2=0.723$, respectively.
Thus (\ref{eq:Glinprec}), which assumes equality between the eigenvalues of $\mat{M}_k^{[t]}\mat{M}_k^{[t]*}$,
does not hold. It can be shown that (let $\gamma_{i}^2=\gamma_{1,\{1,i\}}^2$)
\begin{equation}\label{eq:Ggolden}
\mathcal{G}_{Golden}(\Delta,\Lambda)=
\sqrt{
\left(
\alpha_1\left(\gamma_{1}^2+\gamma_{4}^2\right)+
\alpha_2\left(\gamma_{2}^2+\gamma_{3}^2\right)
\right)
\left(
\alpha_1\left(\gamma_{2}^2+\gamma_{3}^2\right)+
\alpha_2\left(\gamma_{1}^2+\gamma_{4}^2\right)
\right)
}\\
\end{equation}
where $\gamma_{i}^2=\gamma_{1,\{1,i\}}^2$.
As $d_{Hmin}$ increases, $(\gamma_{1}^2+\gamma_{4}^2)/(\gamma_{2}^2+\gamma_{3}^2)$ tends to $1$
for any pairwise error probability and $\mathcal{G}_{Golden}(\Delta,\Lambda)\rightarrow\mathcal{G}_{2,1,opt}(\Delta,\Lambda)$:
The error correcting code limits the coding loss due to the non-equal norm of the sub-parts of the Golden code.
As a remark, if $\gamma_{2}^2+\gamma_{3}^2=0$, which is the worst case, the coding loss is $10\log_{10}(\sqrt{\alpha_1\alpha_2/4})=0.5$~dB.

With DNA precoder and ideal interleaving, Assumption \ref{hyp:mimoconv} is satisfied and
the modified Singleton bound on the diversity order can apply.
All results from the field of error correction coding over block-fading channels directly apply
without any modification to the new $1\times sn_r$ SIMO channel with $n_tn_c/s$ independent blocks.

\subsection{The genie method design criterion for full spreading linear precoders ($s'=n_t$)}

A linear precoding design criterion based on the
genie performance optimization at the detector output has been proposed in \cite{Boutros2003-2}.
When a genie gives a perfect information feedback on the $mn_t$ coded bits required in the APP detector computation,
the performance is computed by averaging all the pairwise error probabilities
obtained when changing only one bit out of $mn_t$. Denote $d$ the distance of the BSK. Assume that the BSK
is transmitted on antenna $l$,  the asymptotic expression of the error probability with genie is 
\begin{equation}\label{eq:asympt_BER_dec_ergo}
P_{genie}(\Delta,\Lambda)\eqv{N_{0}\rightarrow 0}
\C{2n_{r}N_\delta-1}{n_rN_\delta}
\prod_{v=1}^{n_\delta}\left(\frac{\delta_v^2}{2N_0}\right)^{-n_r\lambda_v}
\end{equation}
where $\Delta = \{\delta_v\}$ is the set of square-roots of distinct non-null eigenvalues of $d^2{\mat{S}'}_l^{[t]*}{\mat{S}'}_l^{[t]}$ for all $t$,
$\lambda_v$ their frequency and $N_\delta$ their number.
In the best case, there are $s$ non-null eigenvalues and the coding gain is maximized
if they are equal. First, a sufficient condition to have an equality between the eigenvalues of
${\mat{S}'}_l^{[t_1]*}{\mat{S}'}_l^{[t_1]}$ and  ${\mat{S}'}_l^{[t_2]*}{\mat{S}'}_l^{[t_2]}$ is $\|{\mat{S}'}_l^{[t_1]}\|^2=\|{\mat{S}'}_l^{[t_2]}\|^2$.
Then, all eigenvalues of ${\mat{S}'}_l^{[t]*}{\mat{S}'}_l^{[t]}$ are equal if ${\mat{S}'}_l^{[t]}$ is a unitary matrix,
which leads to the following proposition:

\begin{prop}\label{prop:genie}
A linear precoder achieving a diversity order $sn_r$
with maximum coding gain at the detector output must satisfy the following conditions under
perfect iterative APP decoding of the space-time BICM:
\begin{enumerate}
\itemsep 0pt
\item The $n_s$ subparts of the rows in the $s n_t\times s n_t$ precoding matrix have the
same Euclidean norm
\item In each of the $n_s$ subparts, the $s$ subparts (nucleotides) are orthogonal and
have the same Euclidean norm
\end{enumerate}
\end{prop}

Proposition \ref{prop:genie}, which is more intuitive, is equivalent to Proposition \ref{prop:class_prec}
only if $s'=n_t$, i.e., in case of full spreading.

\subsection{Non-full spreading quasi-optimal linear precoder: DNA cyclotomics}

If $s' \neq n_t$, Proposition \ref{prop:genie} is not optimal in terms of maximum
likelihood performance. 
However, we can split the optimization of a $N_t\times N_t$ linear precoder with spreading factor $s$ 
into $n_t/s'$ optimizations of full spreading $N_t'\times N_t'$ linear precoders with $N_t'=s's$. The optimization of
$\mat{S}$ is now done in two steps:
\begin{enumerate}
\item Apply the genie method to design a full spreading $N_t'\times N_t'$ linear precoder for a $s'\times n_r$ MIMO
channel with $n_s$ blocks, satisfying Proposition \ref{prop:genie},
\item Place the non-null sub-parts in $\mat{S}$ as described in Proposition \ref{prop:class_prec}.
\end{enumerate}

Cyclotomic rotations \cite{Boutros1998} provide good performance on ergodic Rayleigh SISO channels
and have the great advantage to exist for any number of complex dimensions.
Moreover, any coefficient has a unity norm which implies that the norm condition of Proposition \ref{prop:genie}
is naturaly satisfied.
We modified the cyclotomic matrices to satisfy the orthogonality condition in the case of full spreading $s=n_tn_s$.
The coefficients of $\mat{S}$ are equal to
\begin{equation}\label{eq:optlinear}
\begin{array}{ll}
\mat{S}_{l,v+(i-1)n_t+(t-1)n_t^2}=\\
\frac{1}{\sqrt{N_t}}\exp\left(2j\pi\left[(l-1)\left(\frac{1}{\Phi^{-1}(2N_t)}+\frac{(t-1)n_t^2+(i-1)n_t+v-1}{N_t}\right)+ (i-1)\left(\frac{1}{\Phi^{-1}(2n_t)}+\frac{v-1}{n_t}\right)\right]\right)
\end{array}
\end{equation}
\noindent
We denote $\mathcal{S}(n_t,n_s,n_tn_s)$ the modified cyclotomic rotation designed for a $n_t\times n_r$ MIMO block-fading channel, assuming
that the precoder experiences $n_s$ channel realizations. 
The last parameter in $\mathcal{S}(n_t,n_s,n_tn_s)$ denotes the spreading factor.\\

To satisfy Proposition \ref{prop:class_prec}, which gives the design criterion for non-full spreading quasi-optimal linear precoders,
we follow the two steps described above.
Following (\ref{eq:optlinear}),
we first construct $\mathcal{S}(s',n_s,s)$
designed for full spreading of a $s'\times n_r$ MIMO block-fading channel
with $n_s$ channel states in each precoded matrix. 
Then, we place $n_t/s'$ times each subpart of $\mathcal{S}(s',n_s,s)$  in the precoding matrix in order to satisfy Proposition \ref{prop:class_prec}
and construct the quasi-optimal linear precoder $\mathcal{S}(n_t,n_s,s)$ for any set of parameters $n_t$, $n_s$ and $s$. 
Its coefficients are equal to
\begin{equation}
\begin{array}{l}
\forall l_2\in[1,n_t/s'], \forall l_1\in[1,ss'],  \forall t\in[1,n_s],  \forall i\in[1,s'],  \forall v\in[1,s'], \\
\mat{S}_{(l_2-1)s's+l_1,v+(l_2-1)n_t/s'+(i-1)n_t+(t-1)s'n_t}=\\
\frac{1}{\sqrt{N'_t}}
\exp\left(2j\pi\left[(l_1-1)\left(\frac{1}{\Phi^{-1}(2N'_t)}+\frac{v-1+(i-1)s'+(t-1){s'}^2}{N'_t}\right)+ (i-1)\left(\frac{1}{\Phi^{-1}(2s')}+\frac{v-1}{s'}\right)\right]\right)\\
\textrm{ and } 0\textrm{ elsewhere.}
\end{array}
\end{equation}

\subsection{Performance of the quasi-optimal precoder with iterative receiver}

We have presented quasi-optimal linear precoders providing good coding gain and
full diversity ML performance under ideal interleaving.
However, the ML decoder of the global Euclidean code does not exist and we process
iterative joint detection and decoding. Proposition \ref{prop:class_prec}
is satisfied by an infinity of matrices, all providing the same ML performance.
Let us consider the performance behavior after the first iteration.
As no a priori information is available at the detector,
errors before decoding are numerous and not necessarily
transmitted on different precoding time periods. Let us consider one precoding time period
and assume that we observe two erroneous bits. If the bits are transmitted on the same
modulation symbol, the Euclidean distance $d_{k}$ changes but this does not affect the linear precoder optimization.
However, if the two bits are placed onto two different rows of $\mat{S}$,
the average performance might be modified 
and interference inside a block and between blocks should be considered.
An optimization of the precoder following the Tarokh criterion should be done, under the conditions
presented in Proposition \ref{prop:class_prec}. Simulation results show that the modified cyclotomic
rotation has good uncoded ML performance, close to algebraic full rate space-time block codes.
Thus, we expect good performance at the first iteration of a joint detection and decoding process,
which is desirable to reduce the number of iterations needed to achieve the near ML performance
and to provide good performance with non-iterative receivers.
The optimization of the first iteration is not addressed in this paper,
but first answers are given in \cite{kraidy2005-CWIT}.

\section{Practical interleaver design for convolutional 
codes}\label{sec:interleaver}

The maximum diversity to be gathered is limited by the characteristics of the channel,
the linear precoding spreading factor
and the minimum Hamming distance of the binary code, all summarized
in (\ref{equ_upperbounddiv}).
Assume that the linear precoder spreading factor $s$ is chosen 
such that diversity order is maximized, $\Upsilon_{max}=n_c n_t n_r$. 
Thus, there exists an interleaver that allows ML 
performance with full diversity.
We present a new BICM interleaver design 
which satisfies Definition \ref{def:ideal_interleaving} and
leads to the concept of full diversity BICM since the system exhibits a
predetermined diversity whatever the parameters of the considered block-fading 
channel.\\

We first build an interleaver that enables to achieve maximum 
diversity on an $n_t\times n_r$ quasi-static MIMO channel ($n_c=1$) with BPSK input.
Then, we generalize the interleaver construction to apply it to higher spectral efficiency modulations,
linear precoding and finally block-fading channels ($n_c>1$).
%
%
%

\subsection{Interleaver design for quasi-static MIMO channels with BPSK input\label{sec:staticBPSK}}
On quasi-static channels, a codeword undergoes only one channel realization.
Let us consider an error event in the code trellis for which $w$ coded bits 
differ from the transmitted codeword.
As all error events are supposed to have a non-zero probability, the interleaver 
should be designed for any of them. Let us ensure the equi-distribution
property that $L_IN_\mathcal{C}$ successive coded bits,
$L_IN_\mathcal{C}$ being the length of an error path with $L_I$ branches,
are transmitted by all the $n_t$ transmit antennas in the same proportion.
To optimize performance, we must also ensure the non-interference of
erroneous bits within the same time period.
In the ML sense, two interfering erroneous bits may either 
degrade the diversity or the coding gain.
When considering the graph representation of our system model in Fig.~\ref{fig:system_model}, 
a time period corresponds to one channel node. Probabilistic messages
on bits should be independent. Practically, bits inside a channel node should be
connected to distant positions in the code trellis.
These conditions lead to a design criterion for quasi-static channels,
well known in the algebraic space-time coding theory as the {\it rank criterion} \cite{Tarokh1998}
and applied here to the BICM interleaver.\\

To design an interleaver with size $L_\mathcal{C}N_\mathcal{C}$ ensuring
that consecutive bits are mapped on different symbol time periods over all the transmit antennas,
we demultiplex the $L_\mathcal{C}N_\mathcal{C}$ 
coded bits into $n_t$ vectors of length
$L_\mathcal{C}N_\mathcal{C}/n_t$. Each of these $n_t$ sub-frames is
separately interleaved and transmitted on a predetermined transmit antenna. However,
the demultiplexing step is not simply processed via the periodical 
selection of every $n_t$ bits. 
Indeed, some error patterns of convolutional codes have periodic structure. This
may result in non-equally distributed erroneous bits on the $n_t$ transmit antennas
and bad coding gain for these error patterns \cite{Gresset2004-Thesis}.
In order to break periodic structures, we apply the following demultiplexing
\begin{equation}
0\leq i<n_t, ~0\leq j<L_\mathcal{C}N_\mathcal{C}/n_t, 
~~~~~\mat{V}_{i}(j) = \mat{V}\left( \mod{(i+j)}{n_t}+jn_t \right)
\end{equation}
where $\mat{V}$ is the codeword to be demultiplexed, $\mat{V}_{i}$ is the $i$-th 
demultiplexed frame.
This ensures the uniform distribution of erroneous bits
over $n_t$ transmit antennas all along the transmitted frame.
Let us now limit the interference of erroneous bits during the same time period. We 
assume that only simple error events occur. 
If the same interleaver is used for all sub-frames, $n_t$ consecutive bits
are in the same position of interleaved sub-frames and we can limit the interference by
sliding each sub-frame by one bit position and transmit all frames serially on their associated antennas.
Yet, this does not guarantee that $L_IN_\mathcal{C}$
successive bits are transmitted over distinct time periods.
To satisfy this strong condition, we use a particular S-random interleaver \cite{Divsalar1995}
with a sliding input separation which guarantees that any $L_I$ 
successive bits in the interleaved frames are 
not transmitted
during the same block of $n_t$ time periods. If we consider that bit position 
$i$ is
placed at position $\Pi_s(i)$ by the interleaver $\Pi_s$, we should have
\begin{equation}
0\leq j<L_\mathcal{C}N_\mathcal{C}/n_t-L_I, ~~0\leq i <L_I,~~~~~ 
\left\lfloor\frac{\Pi_s(j)}{n_t}\right\rfloor\neq 
\left\lfloor\frac{\Pi_s(j+i)}{n_t}\right\rfloor
\end{equation}
Each of the $n_t$ sub-frames $\mat{V}_{i}$ is interleaved into $\mat{V}_{i}$:
\begin{equation}
0\leq i<n_t, ~~0\leq j<L_\mathcal{C}N_\mathcal{C}/n_t,~~~~~
\mat{V}_{i}\left(\Pi_s(j)\right)=\mat{V}_{i}\left(j\right)
\end{equation}
Then, a new sub-frame $\mat{V}'_{i}$ is built from $\mat{V}_{i}$ as follows:
\begin{equation}
0\leq i<n_t,~~0\leq j_1<L_\mathcal{C}N_\mathcal{C}/n_t^2, ~~0\leq j_2<n_t,
~~~~~
\mat{V}'_{i}\left(\mod{(i+j_2)}{n_t}+j_1n_t\right)=\mat{V}_{i}\left(j_2+j_1n_t\right)
\end{equation}
The above construction keeps blocks of $n_t$ bits of $\mat{V}_{i}$ in positions 
corresponding to the same $n_t$  time periods in $\mat{V}'_{i}$, but with 
a cyclic shift of $i$ positions in a block of size $n_t$.

\subsection{Basic interleaver construction}

Let us generalize the interleaver construction to design a basic interleaver
$\mathcal{I}_{N_I,S_I,L_I}$ for $N_I$ channel inputs, a frame size $S_I$ bits
and a separation $L_I$.
We described $\mathcal{I}_{n_t,L_\mathcal{C}N_\mathcal{C},L_I}$ in the previous 
section. For more general system configurations, the basic interleaver 
$\mathcal{I}_{N_I,S_I,L_I}$ will be used in the sequel.\\

\begin{figure}[!ht]
\centering
\includegraphics[width=0.8\columnwidth]{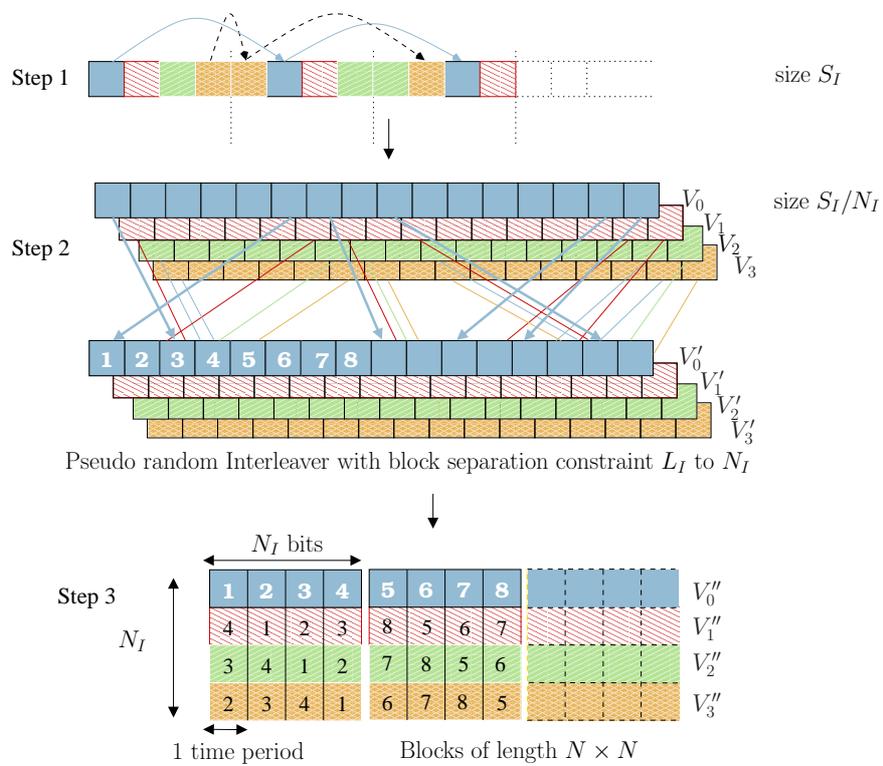}
\caption{Basic interleaver design for $N_I=4$ inputs, a frame size $S_I$ and a 
separation $L_I$}
\label{fig:inter}
\end{figure}

In Fig. \ref{fig:inter}, we present the basic interleaver for $N_I=4$ channel 
inputs. Codeword bits are distinguished by four different patterns,
each pattern corresponding to a specific channel input.
In step 1, the codeword is demultiplexed into $N_I$
sub-frames $\mat{V}_i$, $i=0,\ldots,N_I-1$, of length $S_I/N_I$ each, 
as presented in the previous section.
In step 2, each vector $\mat{V}_i$ of size $S_I/N_I$ is interleaved by the S-random-like 
interleaver into a vector $\mat{V}_i$.
In step 3, we build a $N_I \times S_I/N_I$ matrix as the concatenation of 
$S_I/N_I^2$ matrices of size $N_I\times N_I$. The latter are circulant matrices
where the first row contains the $N_I$ first values of $\mat{V}_0$,
and the second row contains the first $N_I$ values of $\mat{V}_1$.
Rows 3 and 4 are built from $\mat{V}_2$ and $\mat{V}_3$ similarly.

Finally, the $N_I \times S_I/N_I$ matrix is transmitted over the space-time channel 
by distributing its rows on channel inputs and its columns on time periods.\\

This interleaving guarantees that $(L_I-1)N_I+1$ consecutive codeword bits
are not transmitted during the same time period.
The value of $L_I$ of the S-random-like interleaver should be
chosen as large as possible in order to take into account long error events.
An upper bound for $L_I$ can be found based on the interleaver separation
similar to classical S-random \cite{Divsalar1995}.
The interleaver has a sliding input separation equal to $L_I$ and an output
block separation equal to $N_I$ within a sub-frame. Hence, drawing a a simple
two-level tree representation would lead to $(2L_I-1) \times N_I \le S_I/N_I$,
rewritten as
\begin{equation}
L_I \le \frac{1}{2} \left( \frac{S_I}{N_I^2} + 1\right)
\end{equation}
%
%

\subsection{Interleaver design for quasi-static MIMO channels with $M$-ary 
input}
In section \ref{sec:staticBPSK}, we have presented an interleaver for MIMO quasi-static channels and BPSK 
modulation.
For a modulation with higher spectral efficiency, 
erroneous bits in an error path should be dispatched on different time periods 
and equally transmitted over all
the transmit antennas and bit positions. 
Repartition on different bit positions is required as different bits of a modulation
scheme are not equally protected.
These conditions are satisfied by the 
$\mathcal{I}_{mn_t,L_\mathcal{C}N_\mathcal{C},L_I}$ interleaver.

Increasing the diversity by transmitting erroneous bits on all antennas 
is more important than increasing the coding gain by transmitting them on
all modulation bits. 
The $n_t$ first sub-frames should be transmitted on
the $n_t$ transmit antennas and on the first mapping bit. 
The second block of $n_t$ sub-frames should be transmitted on the second mapping bit, and so on.
%
%
\subsection{Application to linear precoding}
When a linear precoder is used to recover a part of the transmit diversity,
the new channel matrix $\mat{S}\mat{H}$ has $sn_t\times sn_t$ rows and columns. 
Linear precoders have been optimized in section \ref{sec:linear-prec}
when at most one erroneous bit is observed on each precoding time period.
We have shown that the precoded channel output is divided into independent
blocks, we modify the order of the rows as follows ($s'=s/n_s$ and 
$N_t'=s's$)
\begin{equation}
\begin{array}{l}
\forall l_2\in[1,n_t/s'], \forall l_1\in[1,ss'],  \forall t\in[1,n_s],  
\forall i\in[1,s'],  \forall v\in[1,s'], \\
\mat{S}_{(l_1-1)n_t/s'+l_2,v+(l_2-1)n_t/s'+(i-1)n_t+(t-1)s'n_t}=\\
\frac{1}{\sqrt{N'_t}}
\exp\left(2j\pi\left[(l_1-1)\left(\frac{1}{\Phi^{-1}(2N'_t)}+\frac{v-1+(i-1)s'+(t-1){s'}^2}{N'_t}\right)+ 
(i-1)\left(\frac{1}{\Phi^{-1}(2s')}+\frac{v-1}{s'}\right)\right]\right)\\
\textrm{ and 0 elsewhere.}
\end{array}
\end{equation}
Now, the $n_t/s'$ consecutive rows of $\mat{S}$ lead to independent row vectors 
$\mat{S}_l\mat{H}_k$ that look like a
true multiple antenna channel.
In this case, the interleaver 
$\mathcal{I}_{smn_t,L_\mathcal{C}N_\mathcal{C},L_I}$ is
designed for diversity and gain exploitation.
As presented in the previous subsection, the $sn_t$ first rows of the last 
interleaver matrix will be transmitted
on the first mapping bit, and so on.

\subsection{Interleaver design for block-fading MIMO channels}
For block-fading channels, $n_c$ different  channel realizations occur during
the codeword. 
In order to take advantage of the transmission and
time diversity given by the linear precoding and the $n_c$ different 
realizations of a block-fading
MIMO channel, the interleaver of a BICM should place consecutive bits on 
different precoding time periods and
equally distribute them among all linear precoding rows and all $n_c$ channel 
realizations.\\
We extract $n_c$ sub-frames from the codeword, each
sub-frame will be transmitted on one of the $n_c$ blocks, and only experience one 
channel realization. We interleave
each sub-frame with the interleaver optimized for MIMO quasi-static channel to 
exploit the linear precoding diversity.\\
The demultiplexing into $n_c$ sub-frames is done in the same manner as for the 
channel inputs in step 1 of Fig. \ref{fig:inter}:
\begin{equation}
0\leq i_{n_c}<n_c, 0\leq j<L_\mathcal{C}N_\mathcal{C}/(n_cn_t), 
\mat{V}_{i}^{n_c}(j)=\mat{V}\left(\mod{(i_{n_c}+j)}{n_c}+jn_c\right)
\end{equation}
This demultiplexing/interleaving is sufficient to exploit the time diversity. 
Indeed, there is no
interference between the symbols experiencing the different channel realizations 
contrary to symbols transmitted
on different linear precoding rows and bit positions.

\subsection{Application to turbo-codes}

The BICM precoder and interleaver have been designed to provide full diversity 
and optimal coding gain for any
pairwise error probability. However, the error rate is given by the 
probability to leave
the Voronoi region. 
With convolutional 
codes,
the number of neighbors increases with the frame length whereas the minimum 
Hamming distance $d_{Hmin}$
remains constant. 
Thus, the frame error rate increases with frame length. 
To obtain the opposite behavior, the Euclidean distance must increase with frame length
and provide a performance gain higher than the performance degradation
due to the increased number of neighbors. 
It has been shown in 
\cite{Guillen2004}\cite{Boutros2004} that
turbo-like codes can fulfill such a condition over block-fading channels.
As proposed in \cite{Gresset2004-Thesis}, we modify the classical parallel turbo code
with two encoders RSC1 and RSC2 and an interleaver $\Pi_t$
by adding a de-interleaving $\Pi_t^{-1}$ of coded bits at the output of RSC2.
Thanks to this de-interleaving, error events are localized
and the optimized channel interleaver can be applied.

\section{Simulation results \label{sec:results}}
In this section, we evaluate the performance of actual iterative joint detection and decoding of the ST-BICM.
The APP detector is performed by exhaustive marginalization.
The set of $2^{mN_t}$ noiseless received precoded symbols $\mat{z}\mat{S}\mat{H}$ is computed once per
channel block realization since the channel matrix $\mat{S}\mat{H}$ is constant during the block. This results 
in a complexity reduction for the marginalization, which now requires around $L_\mathcal{C}N_\mathcal{C}/(msn_t)2^{mN_t}$
operations per iteration if $s \ll L_\mathcal{C}N_\mathcal{C}$.
For large values of $mN_t$, the complexity of the exhaustive search becomes prohibitive.
In order to cope with complexity issues, quasi-optimal or sub-optimal MIMO detectors may also be used,
e.g., a SISO list sphere decoder \cite{Hochwald2003}\cite{Reid2003}\cite{Baro2003}\cite{boutros2003},
a SISO-MMSE detector \cite{Elgamal1999}\cite{Tuchler2002} or a detector using sequential Monte Carlo method \cite{Dong2003}.\\

\begin{figure}[!ht]
   \centering
   \includegraphics[width=0.99\columnwidth]{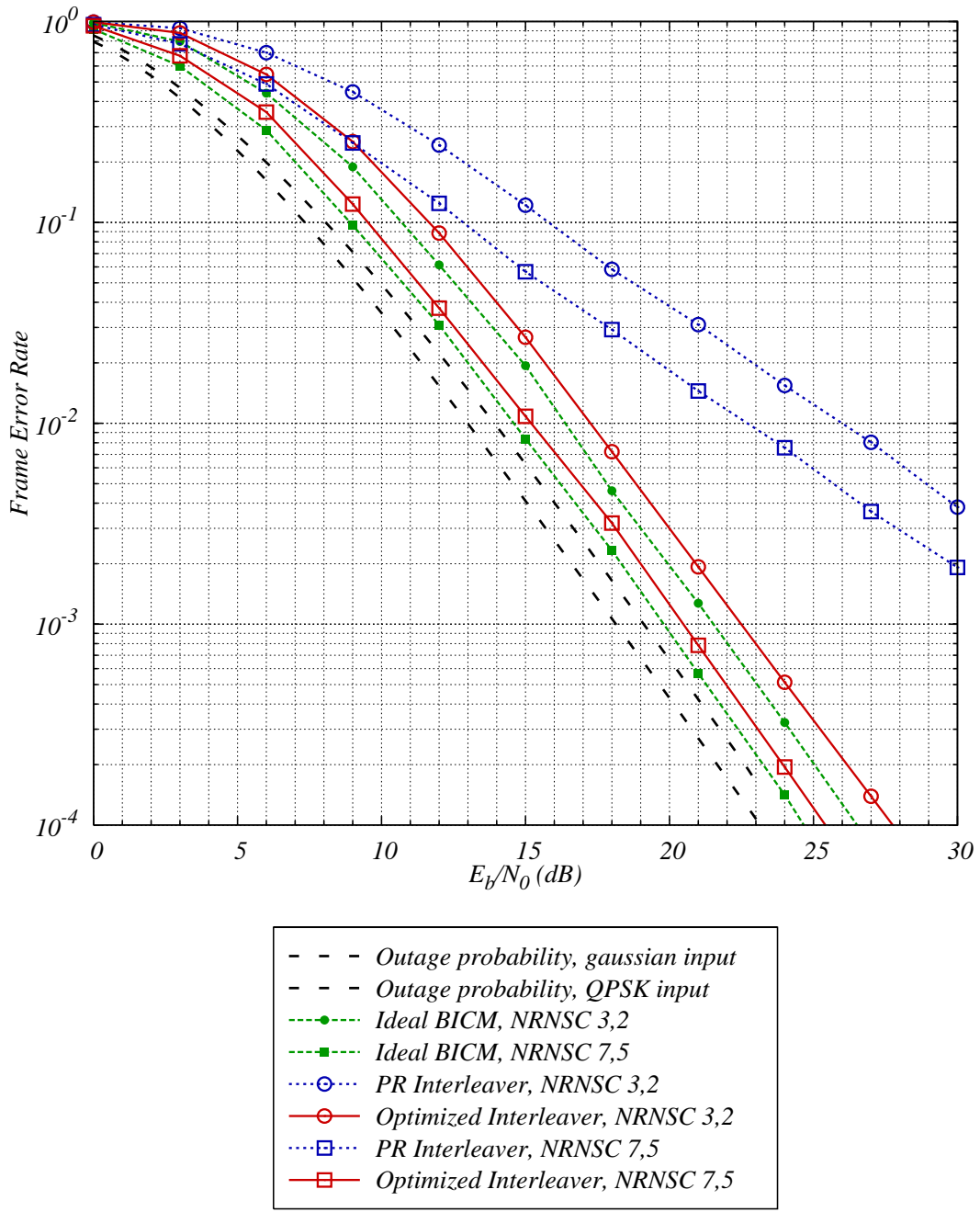}
   \caption{Optimized interleaver with rate-$1/2$ NRNSC codes - QPSK modulation, $2\times 1$ MIMO channel, $n_c=1$, $10$ iterations, $L_cN_c=1024$.}
   \label{fig:compare_inter}
\end{figure}

Let us consider a $2\times 1$ quasi-static ($n_c=1$) MIMO channel and QPSK modulation.
We use $(7,5)_8$ NRNSC or $(3,2)_8$ NRNSC 
codes with rate 1/2 and a blocklength of $1024$ coded bits.
From the Singleton bound we know that full diversity can be achieved without linear precoding.
We compare on Fig. \ref{fig:compare_inter} the performance obtained with a classical PR interleaver
and the performance obtained with the optimized interleaver described in section \ref{sec:interleaver}. 
Full diversity order is only achieved with the optimized interleaver, 
for which the performance slope is equal to the one of the outage probability.
The optimized interleaver provides performance improvement without any
increase of complexity neither at the transmitter nor at the receiver.
In most cases, the PR interleaver only provides a diversity $n_r$, i.e., it does not allow 
any transmit diversity order recovery. The $(7,5)_8$ NRNSC code achieves a higher coding gain than the $(3,2)_8$ NRNSC code.
It achieves performance within only 2.5~dB from the outage capacity with Gaussian input 
and within 1.5~dB from the outage capacity with QPSK input. 
The performance lower bound corresponding to ideally precoded BICM is also drawn.
It is obtained from the performance of the same coded modulation transmitted on a
$1 \times n_cn_tn_r$ SIMO channel, as explained in section \ref{sec:bicm_ideal}.
There is a 1~dB gap between ideal and actual performances with the $(3,2)_8$ NRNSC code and
a 0.75~dB gap with the more powerful $(7,5)_8$ NRNSC code. 
This confirms the analytical result of section \ref{sec:bicm_ideal}
obtained for ML performance: 
The higher the Hamming weight is, 
the closer to the ideal performance the actual iterative receiver can perform.
However, a better code does not always provide better frame error rate.
Indeed, we have seen that, when $w\geq n_tn_c$, 
the full diversity of the considered pairwise error probability can
be achieved with an ideal interleaver. 
The remaining $w-n_tn_c$ BSK distances are uniformly distributed among all the channel states.
A better error correcting code with greater Hamming weights $w'$ does not enhance the diversity
but the coding gain per pairwise error probability.
However, the degradation induced by the increased number of neighbors may be higher than the
improvement brought by increased coding gains. 
How to handle this trade-off is left for further study.\\

\begin{figure}[!ht]
   \centering
   \includegraphics[width=0.99\columnwidth]{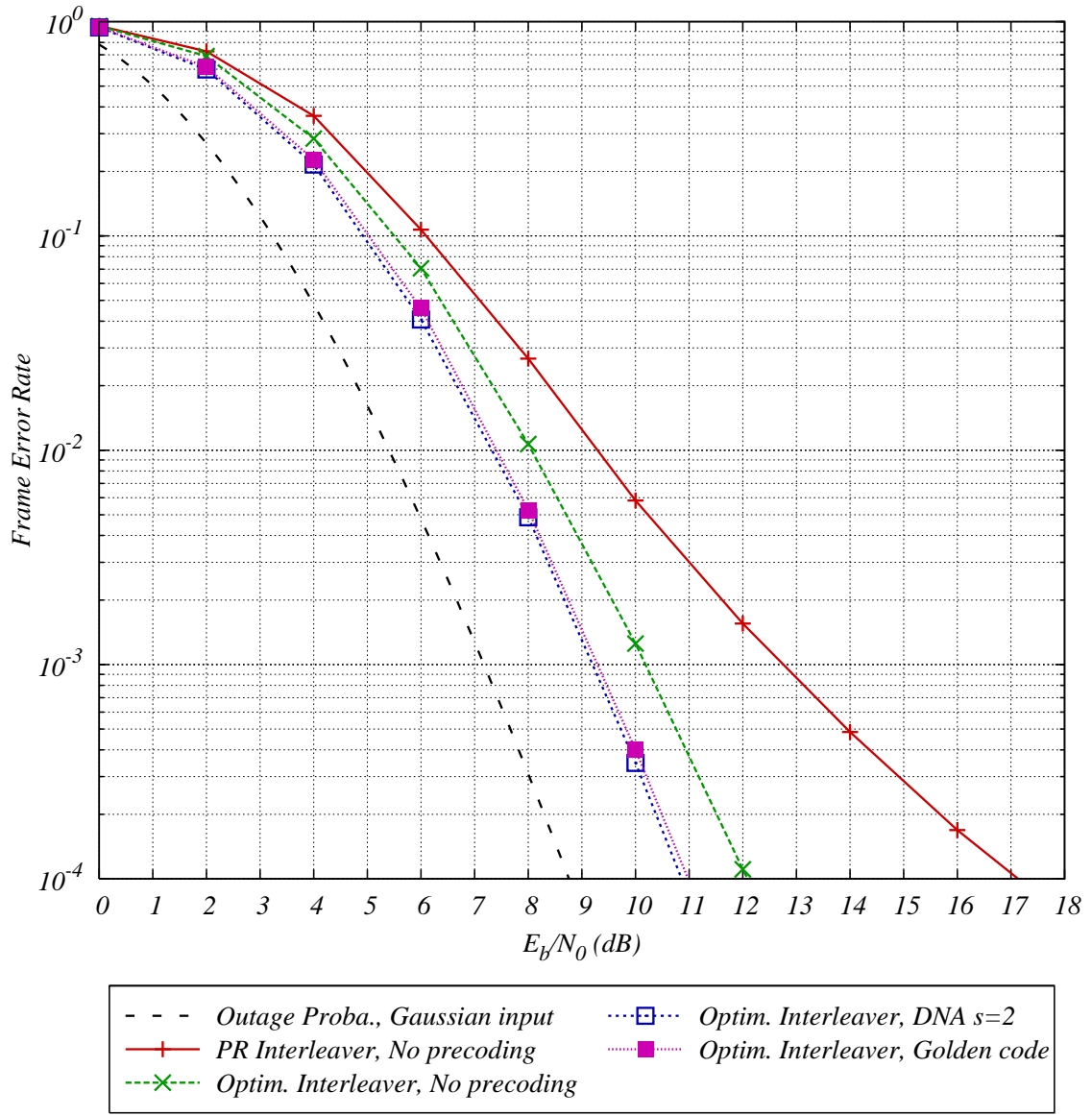}
   \caption{Optimized interleaver with rate-$1/2$ $(7,5)_8$ NRNSC code and linear precoders - 
QPSK, $2\times 2$ MIMO channel, $n_c=2$, $5$ iterations, $L_cN_c=256$ -
No linear precoder,
  DNA cyclotomic precoder ($s=2,n_s=1$), Golden code ($s=2,n_s=1$).}
   \label{fig:St-compare}
\end{figure}

In Fig. \ref{fig:St-compare}, we show the performance of a rate-1/2 $(7,5)_8$ NRNSC code over a $2\times 2$ MIMO block-fading
channel with $n_c=2$ and QPSK input. The frame length is 256 coded bits. 
With a PR interleaver, a diversity order $n_r=2$ is achieved, as transmit diversity is not collected.
Even with the optimized interleaver, full diversity is not obtained at the last iteration.
Indeed, the Singleton bound is equal to $6$ without linear precoding.
Two different linear precoders, the Golden code and the DNA code, both with $s=2$, are used 
to achieve the full diversity order $8$.
The slope difference between diversity orders $6$ and $8$ is not significant.
However, linear precoding provides an additional coding gain which allows 
to perform within 2~dB from the outage capacity with Gaussian input using a
four-state convolutional code and a small frame length. 
The Golden code does not satisfy the equal norm condition, which induces a slight loss in coding gain.
Nevertheless, this loss is fully compensated by the averaging of the $d_{k,l,i}$ into 
equal $\gamma_{k,l}$ values provided by the error correcting code
as explained in \ref{sec:bicm_ideal}.
For a higher frame length, the performance with convolutional codes is degraded.
Therefore, we will also investigate performance with turbo-codes.\\

\begin{figure}[!ht]
   \centering
   \includegraphics[width=0.99\columnwidth]{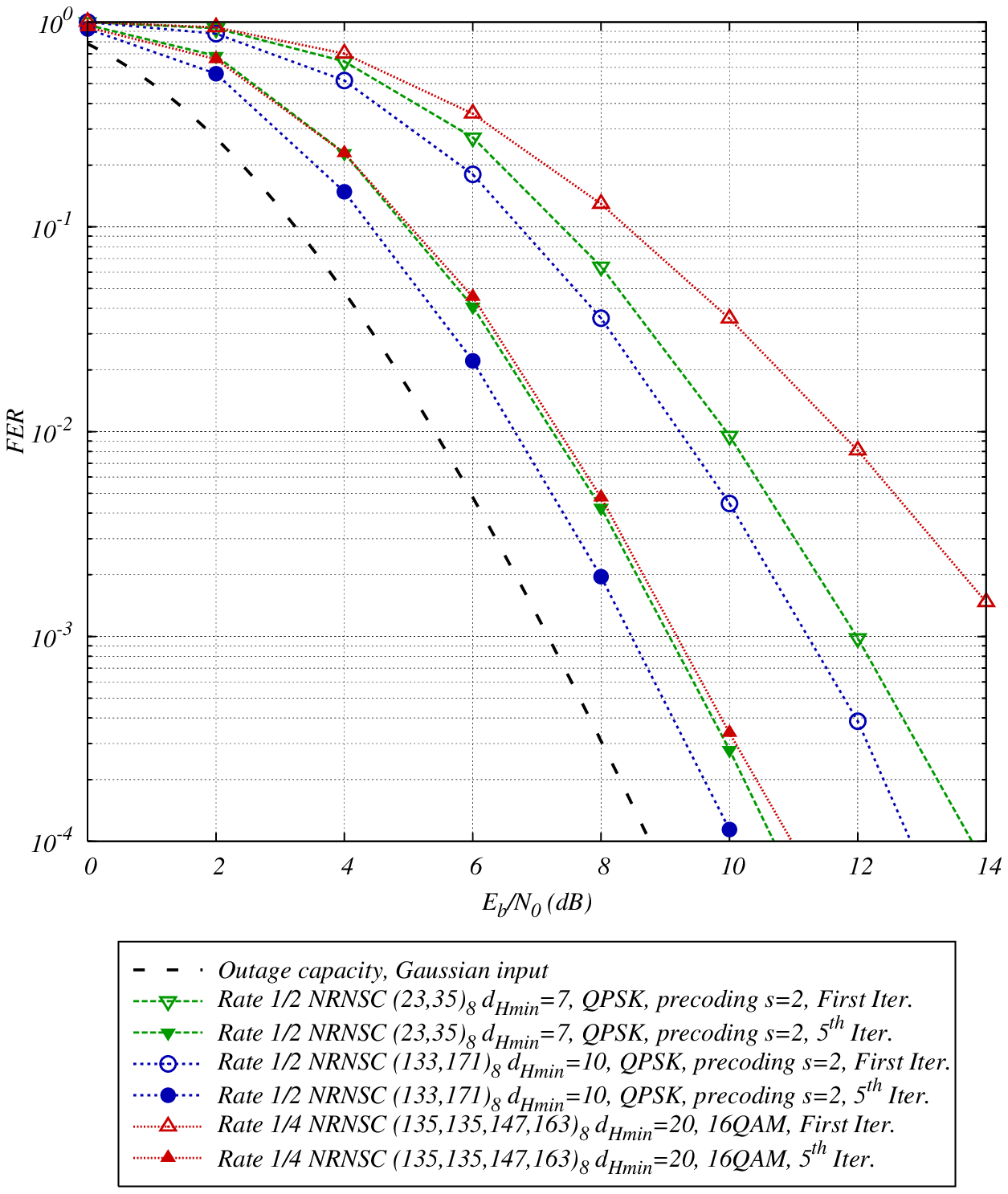}
   \caption{Constellation expansion versus linear precoding - 
$2\times 2$ MIMO channel, $n_c=2$, $L_cN_c=1024$, optimized interleaver.}
   \label{fig:constellexpansion}
\end{figure}

In Fig. \ref{fig:constellexpansion}, we compare two strategies for achieving full diversity with BICM: 
linear precoding and constellation expansion \cite{Guillen2005}. 
Constellation expansion consists in increasing $m$ while decreasing the coding rate,
in order to achieve the full diversity without precoding and with the same spectral efficiency.
A MIMO $2\times2$ channel with $n_c=2$ is considered. The frame length is 1024 coded bits.
Using QPSK modulation and rate-1/2 coding, full diversity is not achieved.
Using a precoded QPSK with $s=2$ and a 16-state rate-1/2 $(23,35)_8$ NRNSC code having
minimal Hamming distance $7$,
we get the same spectral efficiency, 2 bits per channel use, and the Singleton bound is equal to $8$, the full diversity order.
We compare this full-diversity scheme using linear precoding
with a scheme
using constellation expansion from QPSK to 16-QAM with a 64-state rate-1/4 NRCSC code 
having generator polynomials $(135,135,147,163)_8$ and minimal Hamming
distance $20$. With the latter scheme, we get the same spectral efficiency and the Singleton bound is also equal to $8$.
The linear precoder provides a greater diversity order at the first iteration.
At the last iteration, both schemes have same diversity and the precoded scheme slightly
outperforms the scheme with constellation expansion.
Since the detector complexity is around $L_\mathcal{C}N_\mathcal{C}/(msn_t)2^{mN_t}$
operations per iteration if $s \ll L_\mathcal{C}N_\mathcal{C}$, the detection of the precoded system is  
as complex as the detection of the one with constellation expansion. 
However, channel decoding of the 64-state $(135,135,147,163)_8$ 
NRNSC code is more complex than the decoding of the 16-state $(23,35)_8$ NRNSC code.
Thus, to get a same performance, it is less complex to use linear precoding than to use
constellation expansion.
When choosing a 64-state NRNSC $(133,171)_8$ code with rate 1/2 and minimal Hamming distance $10$,
the coding gain is increased by almost 1~dB.\\

\begin{figure}[htb]
   \centering
   \includegraphics[width=0.99\columnwidth]{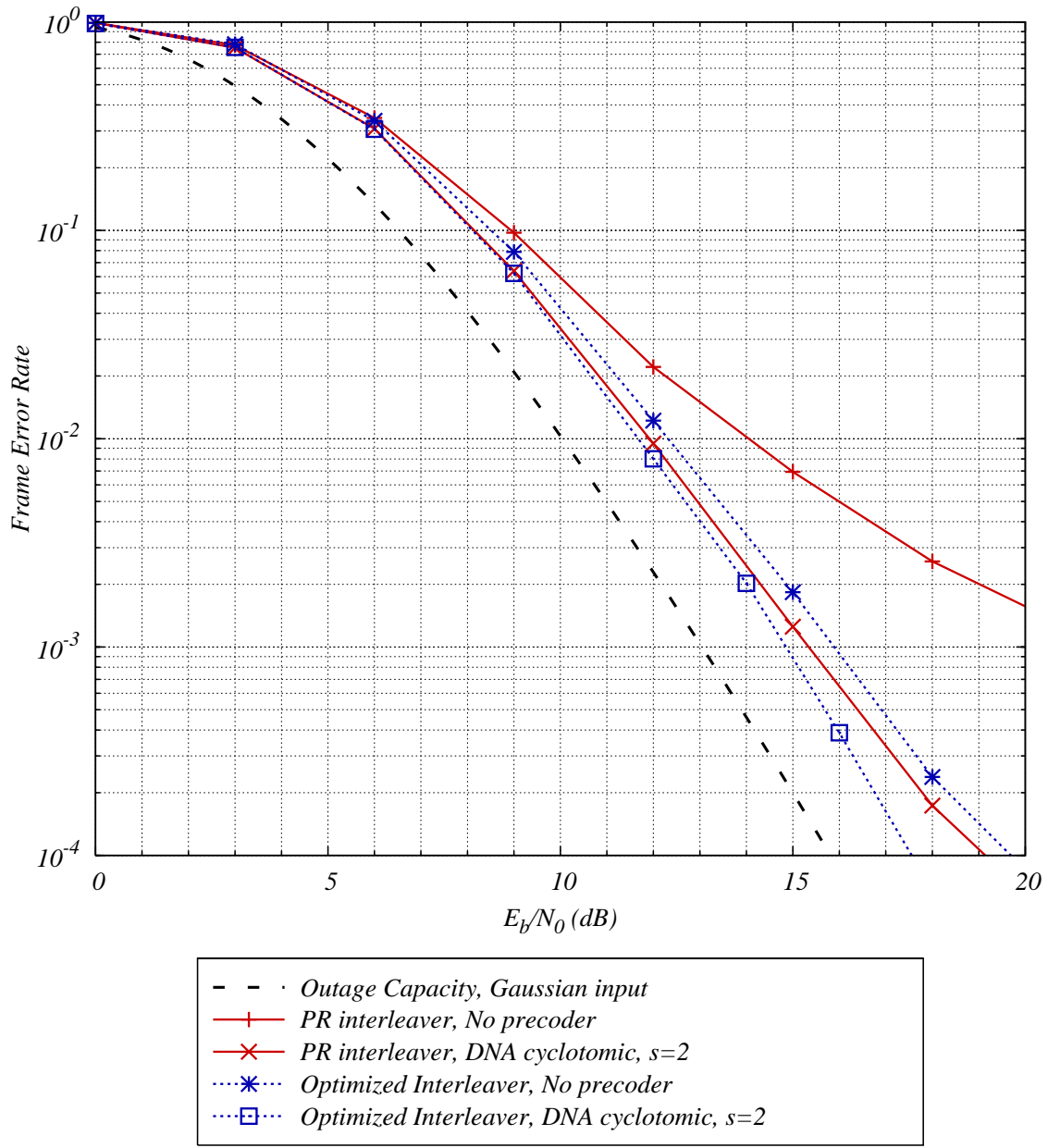}
   \caption{Optimized interleaver with rate-$1/2$ RSC $(7,5)_8$ turbo-code and DNA cyclotomic precoder - 
16-QAM, $1\times 1$ MIMO channel, $n_c=4$, $15$ iterations, $L_cN_c=2048$ -
   Parity check bits of the second constituent are multiplexed via the inverse turbo interleaver.}
   \label{fig:turbo_1x1_4}
\end{figure}

In order to increase the frame length without degrading performance,
we now consider turbo-codes.
Fig. \ref{fig:turbo_1x1_4} illustrates the performance of a $(7,5)_8$ RSC turbo-code over a $1\times1$
channel with $n_c=4$, 16-QAM input and either a PR or an optimized interleaver.
Two different frame lengths ($256$ and $2048$ coded bits) are tested.
With the PR interleaver and without precoding, the full diversity order $4$ is not achieved.
If the optimized interleaver is used, the full diversity order is not achieved neither, but the smaller slope 
is not visible down to a FER equal to $10^{-3}$. 
A similar behavior is obtained with PR interleaver and precoding $s=2$.
Finally, the DNA precoded modulation with optimized interleaver 
achieves full diversity performance within less than 2~dB from
the outage capacity with Gaussian input.\\

\begin{figure}[htb]
   \centering
   \includegraphics[width=0.99\columnwidth]{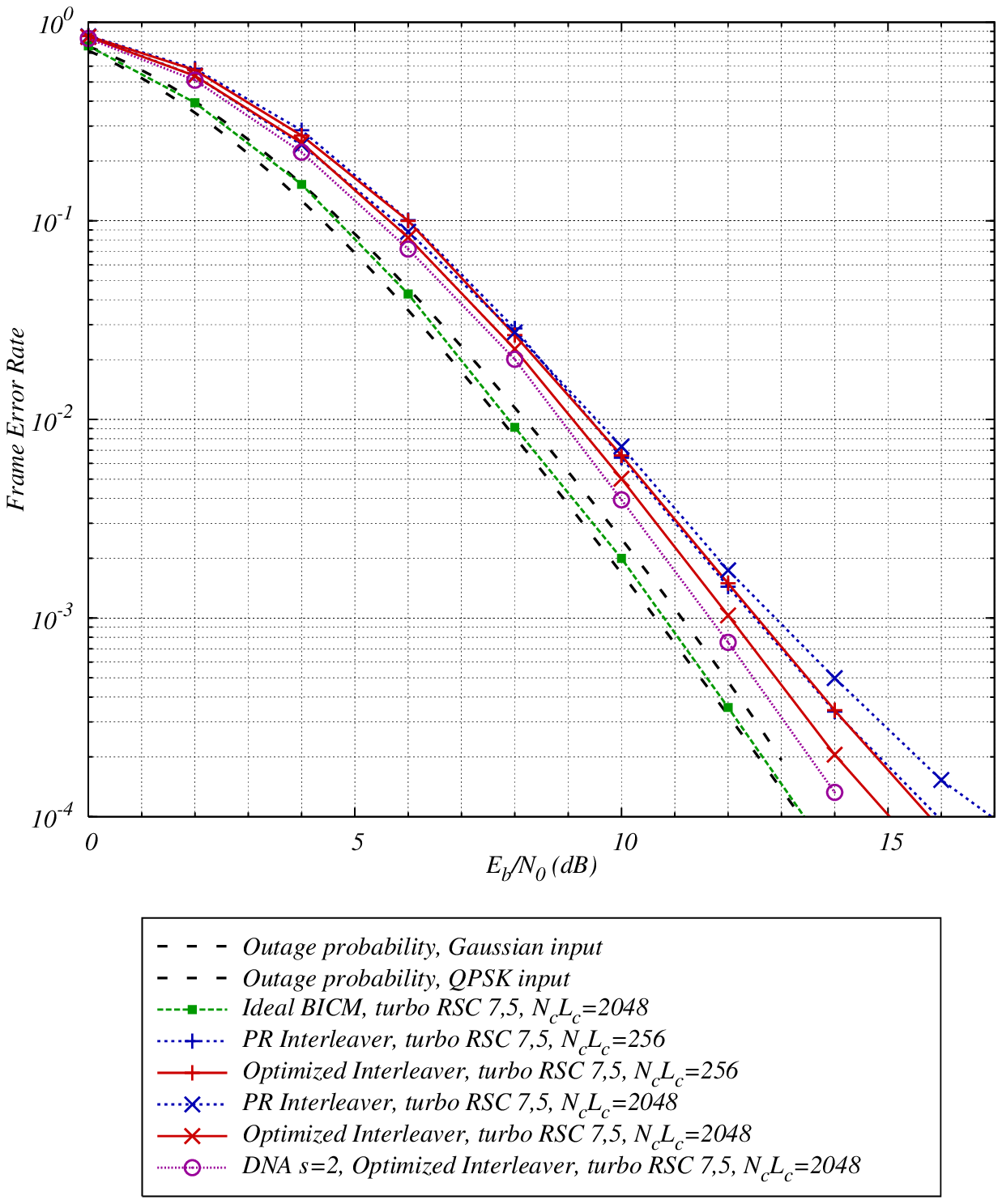}
   \caption{Impact of frame size with a rate-$1/2$ RSC $(7,5)_8$ turbo-code - QPSK, $2\times 2$ MIMO channel, $n_c=1$, $15$ iterations -
   Parity check bits of the second constituent are multiplexed via the inverse turbo interleaver.}
   \label{fig:turbo-2x2}
\end{figure}

Fig. \ref{fig:turbo-2x2} illustrates the performance of a $(7,5)_8$ RSC turbo-code over a $2\times2$ quasi-static
channel with QPSK input and either a PR or an optimized interleaver. 
Two different frame lengths ($256$ and $2048$ coded bits) are tested.
With the PR interleaver, the full diversity order $4$ is not achieved, and the performance degrades when 
the frame length increases, as with convolutional codes. With the optimized interleaver, the full diversity order
is achieved and the frame error rate decreases when the frame length increases. 
The system using DNA precoding ($s=2$), optimized
interleaver and a turbo code finally performs
within 1~dB from the outage capacity with Gaussian input.\\

\begin{figure}[htb]
   \centering
   \includegraphics[width=0.99\columnwidth]{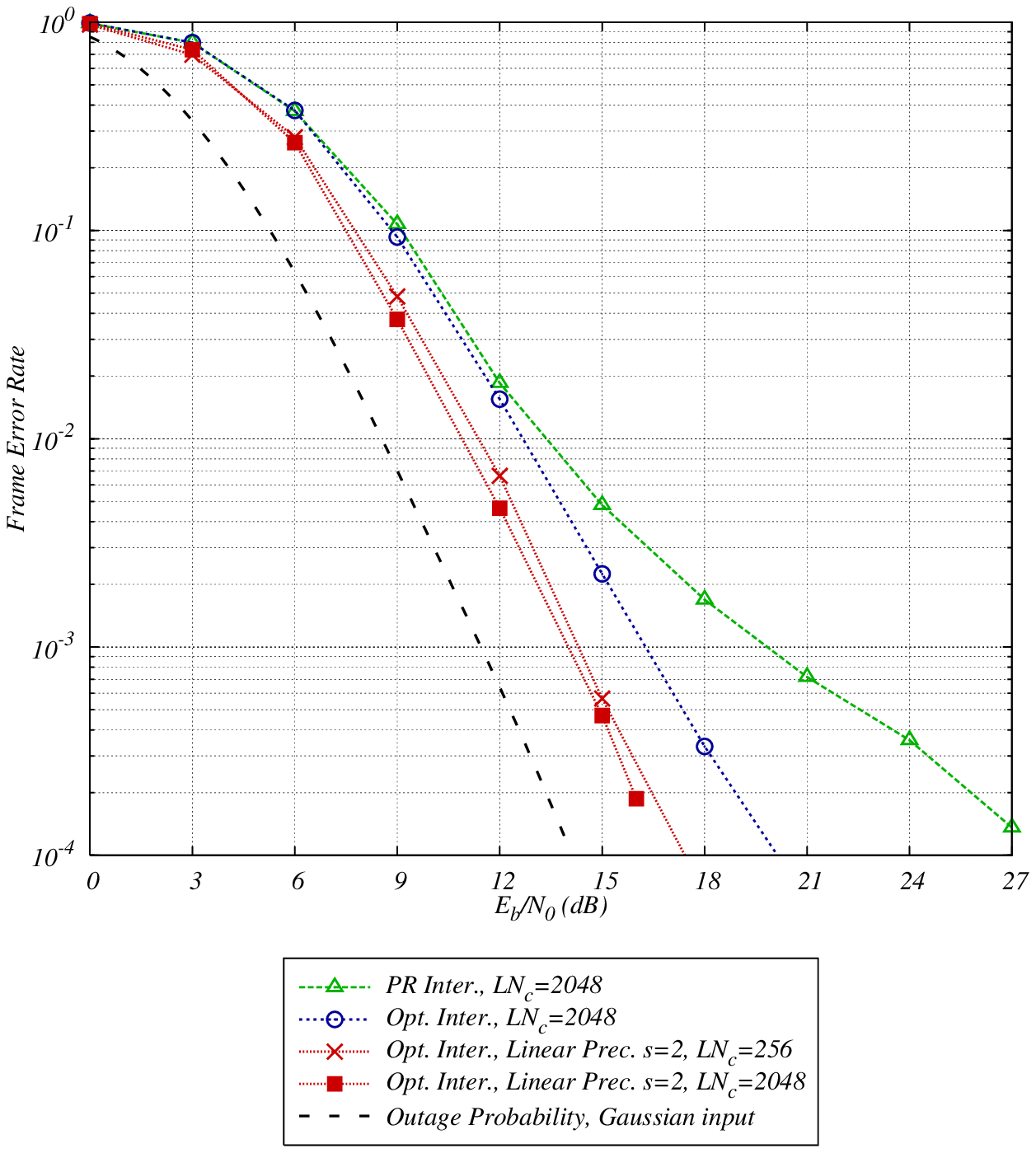}
   \caption{Impact of frame length with a rate-$1/2$ RSC $(7,5)_8$ turbo-code - 
BPSK, $4\times 1$ MIMO channel, $n_c=1$, $15$ iterations -
   Parity check bits of the second constituent are multiplexed via the inverse turbo interleaver.}
   \label{fig:turbo_4x1}
\end{figure}

Fig. \ref{fig:turbo_4x1} represents the performance of a $(7,5)_8$ RSC turbo-code over a $4\times1$ quasi-static
channel with BPSK input and either a PR or an optimized interleaver. 
Two different frame lengths ($256$ and $2048$
coded bits) are tested.
Without linear precoder and using a PR interleaver, the full diversity gain is not achieved.
Asymptotically, the observed diversity is $n_r=1$, but, for low $E_b/N_0$, the performance is close
to the performance obtained with the optimized interleaver. 
Indeed, the turbo-code generates a large amount of errors for low $E_b/N_0$
and the probability of satisfying the ideal interleaving condition with a PR interleaver is
high. However, when $E_b/N_0$ is high, only neighbors have
an influence on the error rate and it is crucial
to place the few erroneous bits on all the channel states. 
This behavior is stressed with increased frame length.
To achieve maximum diversity, according to the Singleton bound, a precoding with at least $s=2$
is needed. This is confirmed
by the simulation results and again the error rate decreases when the frame length increases.
With the $4\times1$ MIMO channel, a large amount of interference exists between the transmit antennas.
Nevertheless, performance is within $2.5$~dB from the outage probability with Gaussian input. 
Performance will be even closer to the outage probability with a higher number of receive antennas
or channel realizations.\\

\begin{figure}[htb]
   \centering
   \includegraphics[width=0.99\columnwidth]{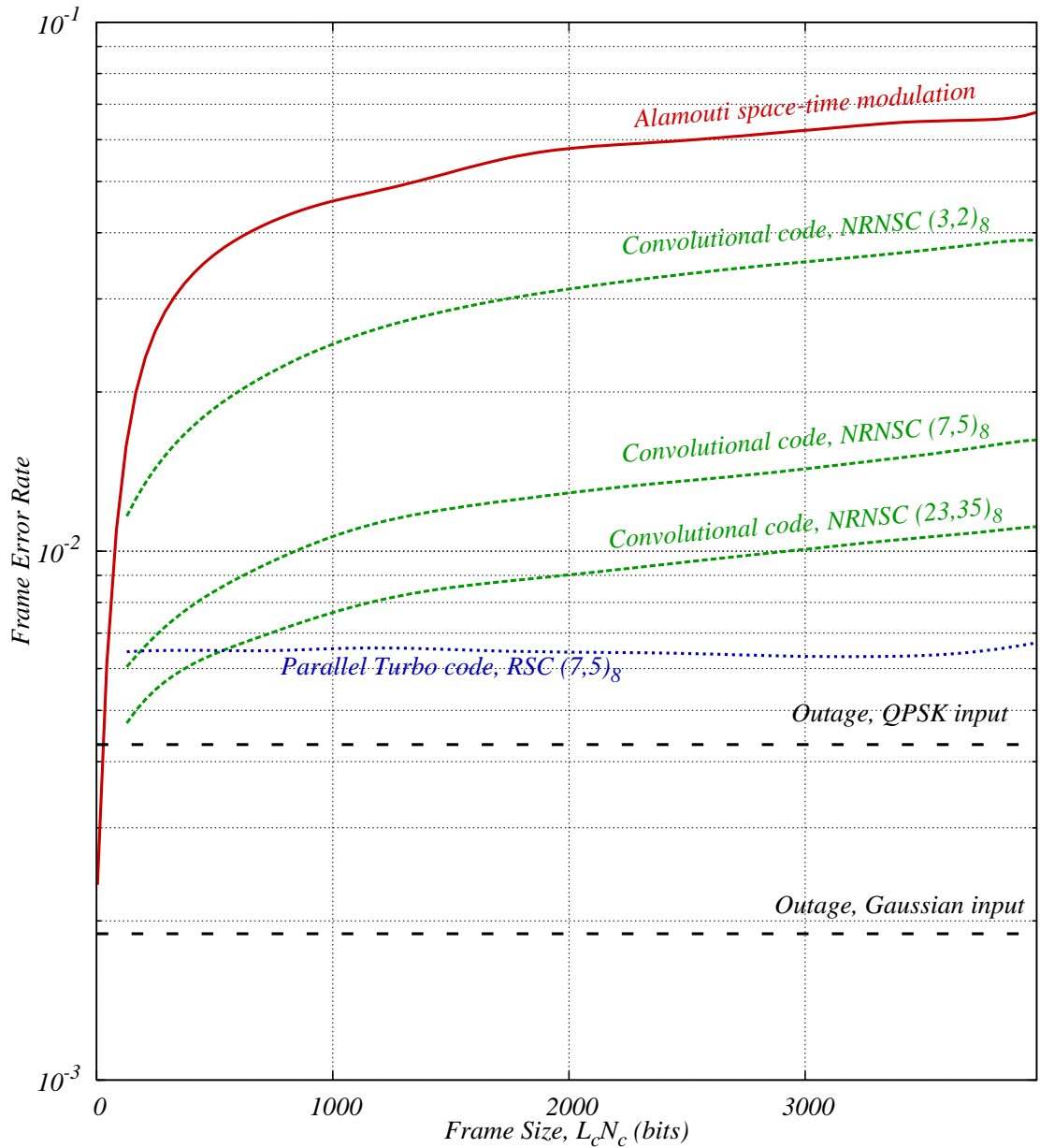}
   \caption{Frame error rate versus the frame size $L_cN_c$ - BPSK, $2 \times 1$ MIMO channel, $n_c=1$, $E_b/N_0$ = 15 dB - Alamouti STBC, NRNSC codes, Rate one half punctured
parallel turbo codes -
   Parity check bits of the second constituent are multiplexed via the inverse turbo interleaver.}
   \label{fig:size2x1}
\end{figure}

On Fig. \ref{fig:size2x1}, performances of NRNSC codes and parallel turbo-codes
with RSC constituent codes  over a $2\times 1$ quasi-static MIMO channel are drawn versus frame size
for $E_b/N_0 = 15$~dB.
Performance of the Alamouti scheme \cite{Alamouti1998} having same spectral efficiency
without channel coding is also drawn as a reference.
The frame error rate increases with the frame size when using Alamouti scheme
or NRNSC codes whereas it remains constant when using turbo codes. 
This strong property may be in part explained by the interleaving
gain of the turbo-code but further research is required on this point. \\

\section{Conclusions}
\begin{figure}[htb]
   \centering
   \includegraphics[width=0.99\columnwidth]{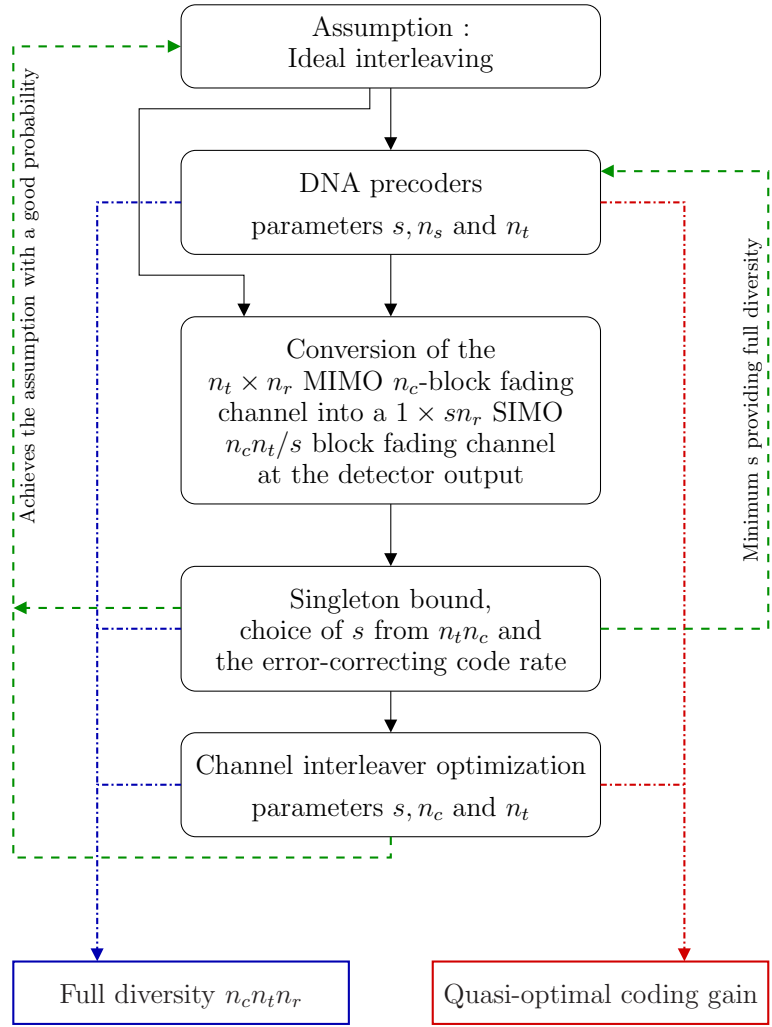}
   \caption{Summary of the space-time BICM optimization process for a 
$n_t\times n_r$ MIMO $n_c$-block fading channel.
   The parameters $s$ and $n_s$ are  the time dimension of the $sn_t\times sn_t$ precoding matrix $S$ and the number
   of independent block channel realizations linked by the precoder respectively.}
   \label{fig:summary}
\end{figure}

In this paper, we have analyzed the ideal behavior of an ST-BICM
using full-rate linear precoding on a MIMO block-fading channel.
Ideal performance has been derived analytically using exact pairwise error probabilities
under ideal interleaving conditions.
Using a bound on the diversity order, we have shown how to set the time dimension
of the linear precoder. Then, we have presented how to design the linear precoder
and the interleaver to obtain an ST-BICM achieving full-diversity
and performing close to the ideal performance and the outage probability.
Fig.~\ref{fig:summary} summarizes the optimization steps followed in this paper.
The proposed DNA precoder slightly outperforms the algebraic Golden code.
Furthermore, the design of DNA precoders holds for any parameter set $(n_t, s, n_s)$,
whereas algebraic codes have to be specifically designed for each pair $(n_t, s)$.
We have also shown that, for a same performance, using linear precoding
is less complex than using constellation expansion.
Finally, using turbo codes with the optimized interleaver, we have obtained
an FER which does not increase with the frame length.

\begin{appendix}
\section{Derivation of $E_{\mat{H}_k}\left[\Psi_{\LLR_{k}}(j\nu)\right]$ for block-fading channels\label{annex:A}}
We first consider $n_s=1$ and extend the result to any value of $n_s$.

\subsection{Precoding matrix experiences one channel realization ($n_s=1$) \label{sss:one_block}}

For $n_s=1$, the quasi-static channel matrix $\mat{H}_k$ is defined as
$\mat{H}_k=\diag\left(\mat{H}_{k}^{[1]}, \ldots, \mat{H}_{k}^{[1]}\right)$, $\mat{H}_{k}^{[1]}$
being repeated $s$ times.
From $\mat{S}_l$,
we construct the $s\times n_t$ matrix 
$\mat{S}'^{[1]}_l=\left( \mat{S}_{l}^{[1][1]~T}, \mat{S}_{l}^{[1][2]~T}, \ldots, \mat{S}_{l}^{[1][s]~T} \right) ^T$.
The row vector $\mat{S}_{l}^{[1][i]}$ of size $n_t$ denotes the $i$-th sub-part of $\mat{S}_l$.
The $n_r$ columns $\mat{h}_{i}$ of $\mat{H}_{k}^{[1]}$ are independent realizations of an $n_t\times 1$
multiple-input single-output channel.
Let us define $\mat{M}^{[1]}_k$ as an $n_t\times n_t$ Hermitian square root matrix of
$\matgr{\Sigma}^{[1]}_k=\sum_{l=1}^{N_t}\gamma_{k,l}^2{\mat{S}'}_{l}^{[1]*}{\mat{S}'}_{l}^{[1]}$. Thus,
\begin{equation}
\mat{M}_k^{[1]}=\mat{M}_k^{[1]*}=U^*\sqrt{\Phi_{\matgr{\Sigma}_k^{[1]}}}U
\end{equation}
where $\Phi_{\matgr{\Sigma}_k^{[1]}}=\diag(\vartheta_{k,1}^{[1]},\ldots,\vartheta_{k,n_t}^{[1]})$, 
$\vartheta_{k,u}^{[1]}$ being the $u$-th real eigenvalue of $\matgr{\Sigma}_k^{[1]}$, 
and $\mat{U}$ is a unitary matrix. We write
\begin{equation}
\sum_{l=1}^{N_t}\mat{R}_{k,l}
=\sum_{i=1}^{n_r}\sum_{l=1}^{N_t} \gamma_{k,l}^2\mat{h}_i^*{\mat{S}'}_{l}^{[1]*}{\mat{S}'}_{l}^{[1]}\mat{h}_i
=\tr\left(\sum_{i=1}^{n_r}\mat{M}_k^{[1]}\mat{h}_i\mat{h}_i^*\mat{M}_k^{[1]*}\right)
\end{equation}
The random variable $\sum_{i=1}^{n_r}\mat{M}_k^{[1]}\mat{h}_i\mat{h}_i^*\mat{M}_k^{[1]}$ has a Wishart distribution with $n_r$ degrees of freedom
and parameter matrix $\matgr{\Sigma}_k^{[1]}$.
The characteristic function of the trace of $\sum_{i=1}^{n_r}\mat{M}_k^{[1]}\mat{h}_i\mat{h}_i^*\mat{M}_k^{[1]}$ is
given in \cite{Maiwald2000}. Finally,
\begin{eqnarray}
E_{\mat{H}_k}\left[\Psi_{\LLR_{k}}(j\nu)\right]
&=&E_{\mat{H}_k}\left[ \exp\left(\frac{\nu(j-\nu)}{2}\frac{\tr(\sum_{i=1}^{n_r}\mat{M}_k^{[1]}\mat{h}_i\mat{h}_i^*\mat{M}_k^{[1]})}{N_0}\right)\right]\\
&=&\left({\det(\matgr{\Sigma}_k^{[1]})\det\left({\matgr{\Sigma}_k^{[1]}}^{-1}-\frac{\nu(j-\nu)}{2N_0}I_{n_t}\right)}\right)^{-n_r}\\
&=&\prod_{u=1}^{n_t}\left(1-\frac{\nu(j-\nu)}{2N_0}\vartheta_{k,u}^{[1]}\right)^{-n_r}\label{eq:psillr_prec_qs}
\end{eqnarray}

\subsection{Precoding matrix experiences several channel realizations ($n_s>1$)}
For $n_s>1$, we first decompose each row $\mat{S}_l$ into $n_s$ sub-parts of size $N_t/n_s$, 
denoted $\mat{S}^{[t]}_l$.
Then, each sub-part $\mat{S}^{[t]}_l$ is decomposed into $s/n_s$ sub-parts $\mat{S}^{[t][i]}_l$ of size $n_t$.
As different values of $t$ correspond to independent channel matrices $\mat{H}_{k}^{[t]}$, 
the characteristic functions associated with the sub-parts $S^{[t]}_l$ can be multiplied.
Substituting $s$ with $s/n_s$ in the mathematical development presented in section \ref{sss:one_block},
we directly have
\begin{eqnarray}
E_{\mat{H}_k}\left[\Psi_{\LLR_{k}}(j\nu)\right]=\prod_{t=1}^{n_s}\prod_{u=1}^{n_t}\left(1-\frac{\nu(j-\nu)}{2N_0}\vartheta_{k,u}^{[t]}\right)^{-n_r}
\end{eqnarray}
where $\vartheta_{k,u}^{[t]}$ is the $u$-th eigenvalue of
\begin{equation} 
\matgr{\Sigma}_k^{[t]}=
\sum_{l=1}^{N_t}\gamma_{k,l}^2\sum_{i=1}^{s/n_s}{\mat{S}}_{l}^{[t][i]*}{\mat{S}}_{l}^{[t][i]}
\end{equation}

\end{appendix}



\newpage
\begin{table}[ht]
\begin{center}
\begin{tabular}[c]{|c|cccccccc|}
\hline
$n_t\diagdown s$&$\:1\:$&$\:2\:$&$\:3\:$&$\:4\:$&$\:5\:$&$\:6\:$&$\:7\:$&$\:8\:$\\
\hline
1&\textbf{1}&&&&&&&\\
2&\textbf{2}&\textbf{2}&&&&&&\\
3&2&&\textbf{3}&&&&&\\
4&3&\textbf{4}&&\textbf{4}&&&&\\
5&3&&&&\textbf{5}&&&\\
6&4&4&\textbf{6}&&&\textbf{6}&&\\
7&4&&&&&&\textbf{7}&\\
8&5&6&&\textbf{8}&&&&\textbf{8}\\
\hline
\end{tabular}
\end{center}
\caption{Diversity order from modified Singleton bound versus number
 of transmit antennas $n_t$ and spreading factor $s$, for $R_\mathcal{C}=1/2$,
 $n_r=1$ and $n_c=1$.\label{tab:table1}}
\end{table}

\begin{table}[ht]
\begin{center}
\begin{tabular}[c]{|c|cccccccc|}
\hline
$n_t\diagdown s$&$\:1\:$&$\:2\:$&$\:3\:$&$\:4\:$&$\:5\:$&$\:6\:$&$\:7\:$&$\:8\:$\\
\hline
1&\textbf{2}&\textbf{2}&&&&&&\\
2&3&\textbf{4}&&\textbf{4}&&&&\\
3&4&4&\textbf{6}&&&\textbf{6}&&\\
4&5&6&&\textbf{8}&&&&\textbf{8}\\
5&6&6&&&\textbf{10}&&&\\
6&7&8&9&&&\textbf{12}&&\\
7&8&8&&&&&\textbf{14}&\\
8&9&10&&12&&&&\textbf{16}\\
\hline
\end{tabular}
\caption{Diversity order from modified Singleton bound versus number of transmit antennas
 $n_t$ and spreading factor $s$, for $R_\mathcal{C}=1/2$, $n_r=1$, $n_c=2$.\label{tab:table2}}
\end{center}
\end{table}

\begin{table}[ht]
\begin{center}
\begin{tabular}{|c|c|c|c|c|c|c|c|}
\hline                &$w=2$	& $w=3$	& $w=4$	& $w=5$	& $w=6$	& $w=7$	&$w=8$	\\ 
\hline $n_t=2$&0.00		&0.26		&0.00		&0.09		&0.00		&0.05		&0.00		\\ 
\hline $n_t=3$&/			&0.00		&0.25		&0.21		&0.00		&0.08		&0.08		\\ 
\hline $n_t=4$&/			&/			&0.00		&0.22		&0.26		&0.17		&0.00		\\ 
\hline $n_t=5$&/			&/			&/			&0.00		&0.19		&0.26		&0.24		\\ 
\hline $n_t=6$&/			&/			&/			&/			&0.00		&0.17		&0.25		\\ 
\hline $n_t=7$&/			&/			&/			&/			&/			&0.00		&0.15		\\ 
\hline $n_t=8$&/			&/			&/			&/			&/			&/			&0.00		\\ 
\hline 
\end{tabular}
\end{center}
\caption{Best gain in dB to be provided by linear precoding with respect to an unprecoded system,
with ideal interleaving and for a given pair of codewords with Hamming distance $w$ and BPSK input.}
\label{tab:gain}
\end{table}

\newpage


\end{document}